\documentclass[aps,prl,superscriptaddress,amsmath,amssymb,floatfix,reprint]{revtex4-2}
\usepackage{graphicx}
\usepackage{dcolumn}
\usepackage{bm}
\usepackage{physics}
\usepackage{color}
\usepackage[colorlinks=true, citecolor=blue, linkcolor=blue, urlcolor=blue]{hyperref}
\usepackage[T1]{fontenc}
\usepackage{tgtermes}

\graphicspath{ {./}{./figs/} }

\begin{document}

\title{Phonons Drive the Topological Phase Transition in Quasi-One-Dimensional Bi$_4$I$_4$}
%
\author{Wenjie Hu}
\thanks{W. H. and J. G. contributed equally to this work}
\affiliation{Centre for Quantum Physics, Key Laboratory of Advanced Optoelectronic Quantum Architecture and Measurement (MOE), School of Physics, Beijing Institute of Technology, Beijing 100081, China.}
\affiliation{International Center for Quantum Materials, Beijing Institute of Technology, Zhuhai, 519000, China.}

\author{Jiayi Gong}
\thanks{W. H. and J. G. contributed equally to this work}
\affiliation{Centre for Quantum Physics, Key Laboratory of Advanced Optoelectronic Quantum Architecture and Measurement (MOE), School of Physics, Beijing Institute of Technology, Beijing 100081, China.}
\affiliation{College of Physics, Chengdu University of Technology, Chengdu 610059, China}

\author{Yuhui Qiu}
\affiliation{Centre for Quantum Physics, Key Laboratory of Advanced Optoelectronic Quantum Architecture and Measurement (MOE), School of Physics, Beijing Institute of Technology, Beijing 100081, China.}
\affiliation{International Center for Quantum Materials, Beijing Institute of Technology, Zhuhai, 519000, China.}

\author{Lexian Yang}
\affiliation{State Key Laboratory of Low Dimensional Quantum Physics, Department of Physics, Tsinghua University, Beijing 100084, China}

\author{Jin-Jian Zhou}
\email{jjzhou@bit.edu.cn}
\affiliation{Centre for Quantum Physics, Key Laboratory of Advanced Optoelectronic Quantum Architecture and Measurement (MOE), School of Physics, Beijing Institute of Technology, Beijing 100081, China.}
\affiliation{International Center for Quantum Materials, Beijing Institute of Technology, Zhuhai, 519000, China.}

\author{Yugui Yao}
\email{ygyao@bit.edu.cn}
\affiliation{Centre for Quantum Physics, Key Laboratory of Advanced Optoelectronic Quantum Architecture and Measurement (MOE), School of Physics, Beijing Institute of Technology, Beijing 100081, China.}
\affiliation{International Center for Quantum Materials, Beijing Institute of Technology, Zhuhai, 519000, China.}

\date{\today}

\begin{abstract}
Quasi-one-dimensional bismuth halides offer an exceptional platform for exploring diverse topological phases, yet the nature of the room-temperature topological phase transition in Bi$_4$I$_4$ remains unresolved. While theory predicts the high-temperature $\beta$-phase to be a strong topological insulator (TI), experiments observe a weak TI.
Here we resolve this discrepancy by revealing the critical but previously overlooked role of electron-phonon coupling in driving the topological phase transition.
%
Using our newly developed \textit{ab initio} framework for phonon-induced band renormalization, we show that thermal phonons alone drive $\beta$-Bi$_4$I$_4$ from the strong TI predicted by static-lattice calculations to a weak TI above $\sim$180~K.
At temperatures where $\beta$-Bi$_4$I$_4$ is stable, it is a weak TI with calculated surface states closely match experimental results, thereby reconciling theory with experiment.
Our work establishes electron-phonon renormalization as essential for determining topological phases and provides a broadly applicable approach for predicting topological materials at finite temperatures.
\end{abstract}

\maketitle

\label{Introduction}
Topological insulators (TIs) constitute a remarkable class of quantum materials characterized by symmetry-protected metallic boundary states, offering significant potential for low-power spintronics and fault-tolerant quantum computing~\cite{Hasan2010,Qi2011,Bansil2016,Ren2016}.
Among the various topological materials discovered, the bismuth halide family Bi$_4$X$_4$(X=Br,I) stands out as a superior and versatile platform due to its unique quasi-one-dimensional (1D) structure and its ability to host diverse topological phases~\cite{Han2022}.
This family has attracted substantial interest since the prediction of large-gap quantum spin Hall (QSH) states in monolayer Bi$_4$Br$_4$~\cite{Zhou2014}. 
The 1D molecular chain architecture of Bi$_4$X$_4$ gives rise to naturally occurring, atomically sharp edges that are ideal for detecting and exploiting topological edge states~\cite{Zhou2015,Han2023}.
This advantage enabled the groundbreaking experimental observation of room-temperature QSH edge states in Bi$_4$Br$_4$~\cite{Shumiya2022,Yang2022,Peng2021}. 
When these chains stack to form bulk crystals, subtle variations in stacking sequences yield multiple exotic topological phases, including the first experimental realizations of the long-sought weak TI~\cite{Noguchi2019,Huang2021,Zhao2024}, higher-order TI~\cite{Noguchi2021,Zhao2023,Hossain2024,Lefeuvre2025}, and three-dimensional QSH insulator~\cite{Yu2024,Yang2024}. 
Compositional engineering further enriches the topological landscape, producing tunable phases such as the weak TI phase in Bi$_4$Br$_2$I$_2$~\cite{Zhong2023,Noguchi2024} and the coalescence of multiple topological orders in Bi$_4$(Br$_{1-x}$I$_x$)$_4$ systems~\cite{Zhong2025}. \\
\indent
However, a fundamental discrepancy between theory and experiment persists for Bi$_4$I$_4$.
Unlike Bi$_4$Br$_4$, monolayer Bi$_4$I$_4$ is predicted to lie near a topological phase boundary~\cite{Zhou2014}.
Variations in the stacking of these monolayers give rise to two bulk phases: the low-temperature $\alpha$-phase and the high-temperature $\beta$-phase~\cite{Hinostroza2024}.
Advanced \textit{ab initio} calculations using hybrid functionals or GW corrections consistently predict $\beta$-Bi$_4$I$_4$ to be a strong TI with Z$_2$ invariants (1;110)~\cite{Liu2016,Autes2016}, whereas multiple experiments identify it as a weak TI with invariants (0;001)~\cite{Noguchi2019,Huang2021,Zhao2024}.
In stark contrast, the same computational approach accurately captures the higher-order TI nature of $\alpha$-Bi$_4$I$_4$, with calculated surface states agree well with experiments~\cite{Liu2022,Zhao2024}.
This contradiction is particularly puzzling given that $\alpha$- and $\beta$-Bi$_4$I$_4$ differ primarily in their stacking sequence~\cite{Noguchi2021}.
The fact that $\beta$-Bi$_4$I$_4$ is stabilized only at elevated temperatures suggests that finite-temperature effects beyond static-lattice calculations may be essential~\cite{Garate2013,Monserrat2016,Antonius2016,QueralesFlores2020,BrousseauCouture2020,Imai2020,Giustino2010,Antonius2014}. \\
\indent
In this work, we develop an \textit{ab initio} approach to compute phonon-induced band renormalization in $\beta$-Bi$_4$I$_4$. 
We show that thermal phonons alone modify the band gap and induce band inversion as temperature increases, 
transforming $\beta$-Bi$_4$I$_4$ from a strong TI, as predicted in static-lattice calculations, to a weak TI above a critical temperature of $\sim$180~K.
Remarkably, this band inversion is driven by phonon-mediated coupling between band edges and high-energy states far from the gap, indicating that high-energy states could shape low-energy topolopy via phonon-mediated processes.
Furthermore, typically neglected off-diagonal self-energies are critical for reopening the gap after band inversion and stabilizing the weak TI phase. 
At temperatures where $\beta$-Bi$_4$I$_4$ is thermodynamically stable, our calculations predict a weak TI with $Z_2$ invariants of (0;001) and gapless topological surface states (TSS) in excellent agreement with angle-resolved photoemission spectroscopy (ARPES) measurements.
Our work resolves the long-standing discrepancy in $\beta$-Bi$_4$I$_4$ and provides a practical \textit{ab initio} framework to predict topological electronic structures in narrow-gap materials at finite temperature. 

\label{methods}
\textit{Theory and methodology}.---Electronic structure renormalization due to electron-phonon (e-ph) coupling at finite temperature can be calculated using many-body perturbation theory. At the lowest-order, the e-ph self-energy consists of two contributions: the Fan term and the Debye-Waller (DW) term~\cite{Allen1976,Allen1981,Allen1983,Giustino2017,Nery2018}:
\begin{equation} \label{eq:e1}
\Sigma_{n n^{\prime}}(\bm{k},\omega) = \Sigma_{n n^{\prime}}^{\text{Fan}}(\bm{k},\omega) + \Sigma_{n n^{\prime}}^{\text{DW}}(\bm{k}).
\end{equation}
Here, $n, n^{\prime}$ are the band indices, $\bm{k}$ is the crystal momentum. The DW self-energy is static, thus independent of frequency $\omega$. Under the rigid-ion approximation, the DW self-energy including the off-diagonal terms, can be evaluated as~\cite{Ponce2014,Lihm2020}:
\begin{equation}\label{eq:dw}
    \Sigma_{n n^{\prime}}^{\text{DW}}(\bm{k})  =\frac{1}{N_{\bm{q}}}\sum_{ \substack{\kappa \alpha \alpha^{\prime} \\ \nu\bm{q}}} 
    \frac{e_{\nu\bm{q}}^{\kappa \alpha *} e_{\nu\bm{q}}^{\kappa\alpha^{\prime}}}{2\omega_{\nu\bm{q}} M_{\kappa}} \mathcal{D}_{n n^{\prime}}^{\kappa \alpha \alpha^{\prime}}(\bm{k}) \left(n_{\nu\bm{q}}+1/2\right),
\end{equation}
where $\omega_{\nu\bm{q} }$ and $e_{\nu\bm{q} }$ are the phonon energy and eigenvector, respectively, $n_{\nu\bm{q}}$ is the Bose occupation number. $\mathcal{D}_{n n^{\prime}}^{\kappa \alpha \alpha^{\prime}}(\bm{k}) = i\left\langle u_{n \bm{k}}\right|\left[\partial_{\Gamma \kappa \alpha} \hat{v}_{\text{KS}}, \hat{p}_{\alpha^{\prime}}\right] \left|u_{n^{\prime} \bm{k}}\right\rangle$, where $|u_{n\bm{k}}\rangle$ is the periodic part of the electron wavefunction, $\partial_{\Gamma \kappa \alpha} \hat{v}_{\text{KS}}$ is the derivative of the Kohn-Sham (KS) potential $\hat{v}_{\text{KS}}$ with respect to the displacement of atom $\kappa$ (with mass $M_{\kappa}$) in the Cartesian direction $\alpha$ with wavevector $\bm{q}=\Gamma$, and $\hat{p}_{\alpha^{\prime}}$ is the momentum operator.

The Fan self-energy is given by $\bm{\Sigma}^{\text{Fan}} = i\bm{G_{0}}\bm{D_{0}}$, where $\bm{G_{0}}$ and $\bm{D_{0}}$ are the non-interacting electron and phonon propagators in matrix notation, respectively. It can be expressed in terms of electron band energy $\varepsilon_{n \bm{k}}$ and $\omega_{\nu\bm{q} }$ as~\cite{Giustino2017}:
\begin{equation}\label{eq:fan}
  \begin{aligned}
        \Sigma_{n n^{\prime}}^{\text {Fan}}(\bm{k},\omega)  =  & \frac{1}{N_{\bm{q}}}\sum_{m}  \sum_{\nu \bm{q}} g_{m n \nu}(\bm{k},\bm{q})^{*}  g_{m n^{\prime}\nu  }(\bm{k},\bm{q}) 
        \\ & \times \sum_{\pm} \frac{ n_{\nu \bm{q}}+\left[1 \pm\left(2 f_{m \bm{k}+\bm{q}}-1\right)\right] / 2}{ \omega-\varepsilon_{m \bm{k}+\bm{q}} \pm \omega_{\nu \bm{q}}+i \eta} ,
    \end{aligned}
\end{equation}
where $g_{m n\nu}(\bm{k},\bm{q})$ is the e-ph matrix element, $\eta$ a small broadening, and $f_{m \bm{k}+\bm{q}}$ the Fermi occupation number.

The renormalized electronic structure is obtained from the total e-ph self-energy. To properly account for the off-diagonal terms of the self-energy, we employ the static Hermitian approximation, which extends the conventional on-the-mass-shell approximation to the off-diagonal case~\cite{Schilfgaarde2006,Lihm2020}.
Within this approximation, the renormalized electron energies are determined by constructing and diagonalizing the corresponding renormalized Hamiltonian:
\begin{equation} \label{eq:ham}
\bm{\tilde{H}}_{}(\bm{k}) = \bm{H}^{0}(\bm{k}) + \frac{1}{2} \left [ \bm{\tilde{\Sigma}}(\bm{k})  + \bm{\tilde{\Sigma}}^{\dagger}(\bm{k}) \right] ,
\end{equation}
where $H^{0}_{n n^{\prime}}(\bm{k}) = \varepsilon_{n\bm{k}}\delta_{nn^{\prime}}$ and $\tilde{\Sigma}_{n n^{\prime}}(\bm{k})$ is defined as $[\Sigma_{{n n^{\prime}}}(\bm{k}, \varepsilon_{n\bm{k}}) + \Sigma_{{n n^{\prime}}}(\bm{k}, \varepsilon_{n^{\prime}\bm{k}})] / 2$. 

\begin{figure*}[!htbp]
\centering
\includegraphics[width=2\columnwidth]{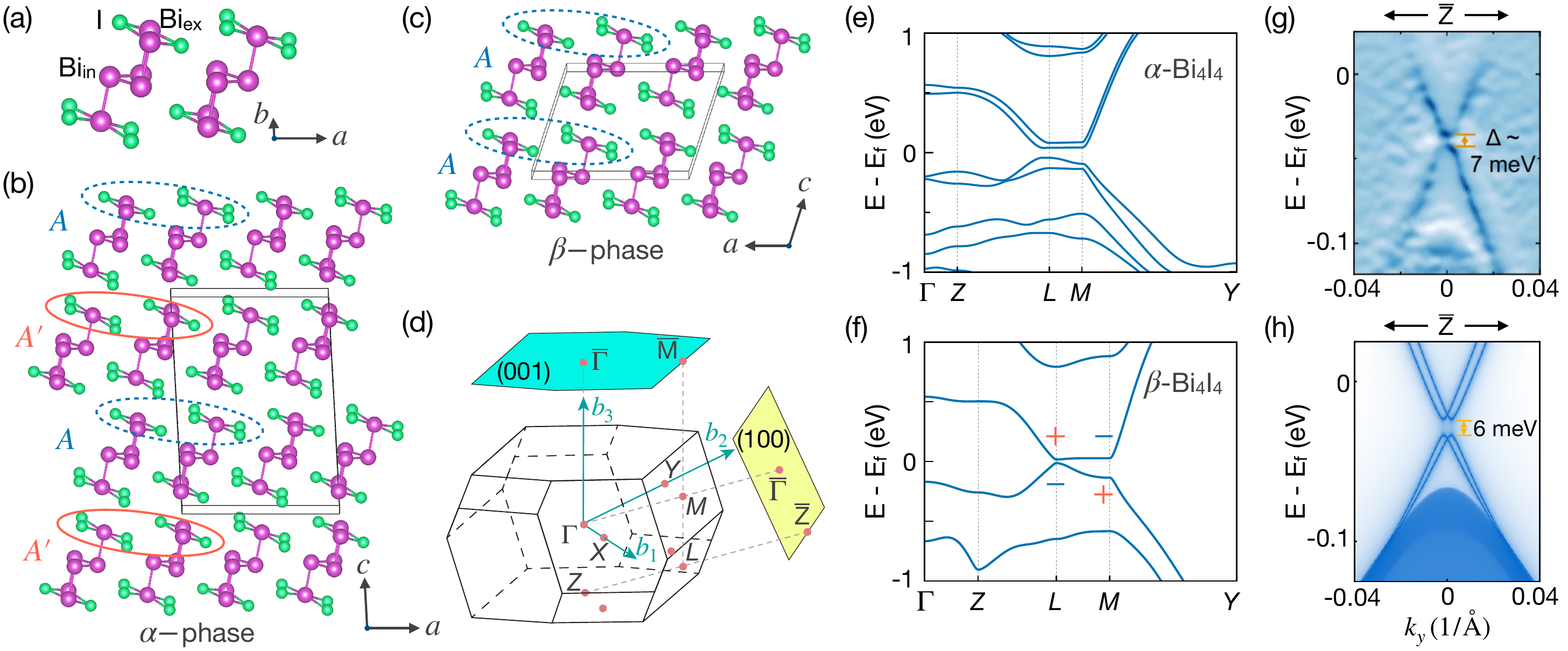}
\caption{(a) Conventional unit cell of monolayer Bi$_4$I$_4$. Crystal structures of (b) $\alpha$-Bi$_4$I$_4$ and (c) $\beta$-Bi$_4$I$_4$. The $A^{\prime}$ layer can be viewed as the $A$ layer shifted by one chain along the $a$-direction. (d) The Brillouin zone of $\beta$-Bi$_4$I$_4$ and its projections onto the (100) and (001) surfaces.  Band structures of (e) $\alpha$-Bi$_4$I$_4$ and (f) $\beta$-Bi$_4$I$_4$ from HSE calculations. The $\pm$ symbols in (f) label the parities of bands at $L$ and $M$ points. (g) ARPES spectrum and (h) calculated surface states of the $\alpha$-Bi$_4$I$_4$ (100) surface. The ARPES spectrum is taken from Ref.~\cite{Zhao2024}.}
\label{Fig:fig1}
\end{figure*}

In practical calculations, the summation over band index $m$ in Eq.~(\ref{eq:fan}) poses a formidable challenge for converging the Fan self-energy~\cite{Ponce2025}, typically requiring thousands of high-energy empty bands. To circumvent this explicit summation, we employ the Sternheimer equation technique~\cite{Gonze2011,Lihm2021}. The full band space is partitioned into two subspaces: a finite set of bands near the band gap or Fermi level (lower subspace) and all remaining high-energy bands (upper subspace). The lower-subspace contribution (lower Fan) is computed directly from Eq.~(\ref{eq:fan}) with $m$ restricted to the lower-space bands. The upper-subspace contribution (upper Fan) is obtained by solving the linear Sternheimer equation, as detailed in the Supplemental Material (SM)~\cite{supmat}.

We calculate the band structures, phonon dispersions, and derivatives of the KS potential $\partial_{\bm{q} \kappa \alpha} \hat{v}_{\text{KS}}$ 
within the Perdew-Burke-Ernzerhof generalized gradient approximation (GGA) of density functional theory (DFT) and density functional perturbation theory (DFPT) using the \textsc{Quantum Espresso} package~\cite{Giannozzi2009,Giannozzi2017}.
These quantities are used to evaluate the DW self-energy~[Eq.~(\ref{eq:dw})], upper-Fan self-energy~[Eq.~(S5) in SM~\cite{supmat}], and e-ph matrix elements on coarse $\bm{k}$- and $\bm{q}$-point grids. 
Wannier interpolation is then employed to obtain e-ph matrix elements on ultrafine grids required to converge the lower-Fan self-energy~\cite{Zhou2021}.
For $\beta$-Bi$_4$I$_4$, we select 48 bands derived from Bi-$p$ and I-$p$ orbitals as the lower subspace (Fig.~S1 in SM~\cite{supmat}) and construct corresponding Wannier functions~\cite{Marzari2012,Pizzi2020}.

To efficiently obtain the renormalized band structure at arbitrary $\bm{k}$, we construct the renormalized Hamiltonian in Wannier basis. We first compute and converge the total e-ph self-energy on a coarse $\bm{k}_c$-grid, yielding the renormalized Hamiltonian in the band basis from Eq.~(\ref{eq:ham}), which is then transformed to the Wannier basis using the same unitary matrix $\bm{\mathcal{U}}$ employed in the Wannier function construction:
\begin{equation} \label{eq:ham_wan}
\bm{\tilde{H}}(\bm{R}) =  \frac{1}{N_{\bm{k}_{c}}} \sum_{\bm{k}_{c}} e^{-i\bm{k}_{c} \cdot \bm{R} } \, \bm{\mathcal{U}}^{\dagger}(\bm{k}_{c}) \bm{\tilde{H}}(\bm{k}_{c})\bm{\mathcal{U}}(\bm{k}_{c}),
\end{equation}
where $\bm{R}$ is the lattice vector. Renormalized band energies within the lower subspace can be efficiently computed from $\bm{\tilde{H}}(\bm{R})$ at negligible cost.
Alternatively, one may directly interpolate the DW and Fan self-energies at arbitrary $\bm{k}$ using Wannier function and its perturbation theory~\cite{Lihm2021}. 
We have verified both approaches yield consistent results. In this work, we adopt the former, which is more efficient and facilitates orbital-projected band analysis and surface-state calculations via the surface Green's function method~\cite{Sancho1985}.

Band structures from DFT at the GGA-level often lack sufficient accuracy, requiring corrections from advanced methods such as GW, which add an electron-electron (e-e) self-energy. 
For $\beta$-Bi$_4$I$_4$, GW corrections or hybrid functionals are required, but DFPT with these methods is computationally prohibitive~\cite{Li2019}.
To address this, we adopt a recently developed scheme that incorporates the e-e self-energy into e-ph self-energy calculations~\cite{Abramovitch2023,Abramovitch2024}.
Specifically, we perform hybrid-functional (HSE) calculations to obtain the electronic Hamiltonian in the same Wannier basis as used in our GGA calculations~\cite{supmat};
the difference between the HSE and GGA Hamiltonians defines an effective e-e self-energy in the Wannier basis, 
which is applied to the Fan self-energy via a corrected electron propagator $\bm{G_{0}}$, and added to the renormalized Hamiltonian.
The above computational framework is implemented in our \textsc{Perturbo} package~\cite{Zhou2021}. Further details on the methodology are provided in the SM~\cite{supmat}.

\label{results-A}
\textit{Static-lattice calculations}.---Monolayer Bi$_4$I$_4$ is composed of parallel 1D molecular chains running along $b$-axis, with adjacent chains offset by $b/2$ [Fig.~\ref{Fig:fig1}(a)]. These monolayers stack along $c$-axis to form bulk crystals via two distinct patterns: the low-temperature $\alpha$-phase adopts the $AA^{\prime}$-stacking with a double-layer unit cell [Fig.~\ref{Fig:fig1}(b)], while the metastable $\beta$-phase exhibits the simpler $A$-stacking with a single-layer unit cell [Fig.~\ref{Fig:fig1}(c)].
$\beta$-Bi$_4$I$_4$ is thermodynamically stable only above the structural transition temperature $T_P\approx300$~K and transforms into the $\alpha$-phase below $T_P$~\cite{Hinostroza2024}. \\
\indent
Before incorporating phonon effects, we establish the baseline band structures using static-lattice HSE calculations~\cite{supmat}, 
which also serve as a starting point for subsequent analysis of e-ph renormalization. Figures~\ref{Fig:fig1}(e) and~\ref{Fig:fig1}(f) present calculated band structures for $\alpha$- and $\beta$-phase, respectively. For $\alpha$-Bi$_4$I$_4$, HSE calculations yield excellent agreement with experiment. 
The computed bulk band gap of $\sim$76~meV closely matches the ARPES-measured $\sim$85~meV~\cite{Huang2021}. 
Calculated surface states [Fig.~\ref{Fig:fig1}(h), also Fig.~S2 in SM~\cite{supmat}] reproduce key features of the ARPES spectrum [Fig.~\ref{Fig:fig1}(g)], including a surface-state gap of 6~meV versus 7~meV in experiment.
Although the calculated Z$_2$ invariants yield (0;000), $\alpha$-Bi$_4$I$_4$ hosts topological hinge states characteristic of a higher-order TI (Fig.~S3 in SM~\cite{supmat}).
This excellent agreement validates our computational approach for the low-temperature phase. \\
\indent
For $\beta$-Bi$_4$I$_4$, the same methodology predicts a bulk gap of $\sim$30~meV and Z$_2$ = (1;110), indicating a strong TI [Fig.~\ref{Fig:fig1}(f)].
This result is consistent with previous calculations using GW corrections~\cite{Autes2016} or hybrid functionals~\cite{Liu2016}.
However, it contradicts experimental evidences pointing to a weak TI~\cite{Noguchi2019,Huang2021,Zhao2024}, 
and the calculated surface states show clear discrepancies with ARPES spectra (see Fig.~S4 in SM~\cite{supmat}).
This selective failure for the high-temperature $\beta$-phase suggests that phonon effects beyond static-lattice calculations may play a critical role.

\begin{figure}[!htbp]
\centering
\includegraphics[width=\columnwidth]{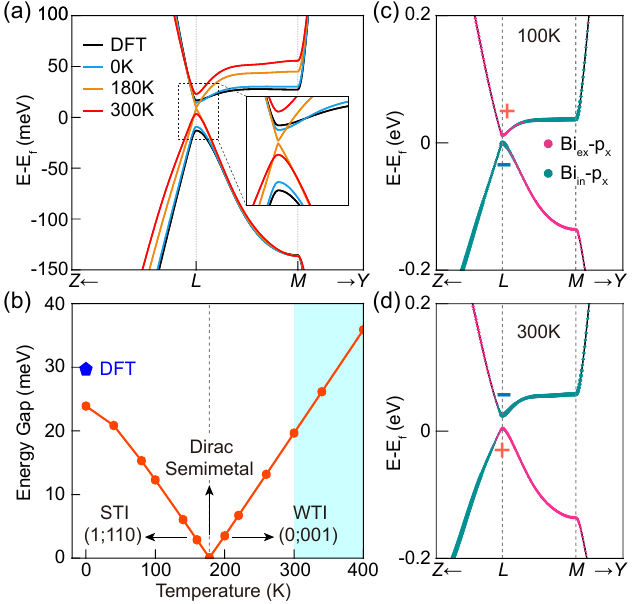}
\caption{(a) Phonon-renormalized band structures of $\beta$-Bi$_4$I$_4$ at different temperatures, with static-lattice DFT result shown for comparison. (b) Temperature-dependent band gap and $Z_2$ topological invariants. The shaded region indicates the temperature range where $\beta$-Bi$_4$I$_4$ is thermodynamically stable. The Bi$_\text{in}$-$p_x$ and Bi$_\text{ex}$-$p_x$ orbital-projected characters of the conduction and valence bands at (c) $T = 100$~K and (d) $T = 300$~K, revealing band inversion near the $L$ point. The $\pm$ symbols label the parities of bands at the $L$ point.}
\label{Fig:fig2}
\end{figure}

\begin{figure}[!htbp]
\centering
\includegraphics[width=0.82\columnwidth]{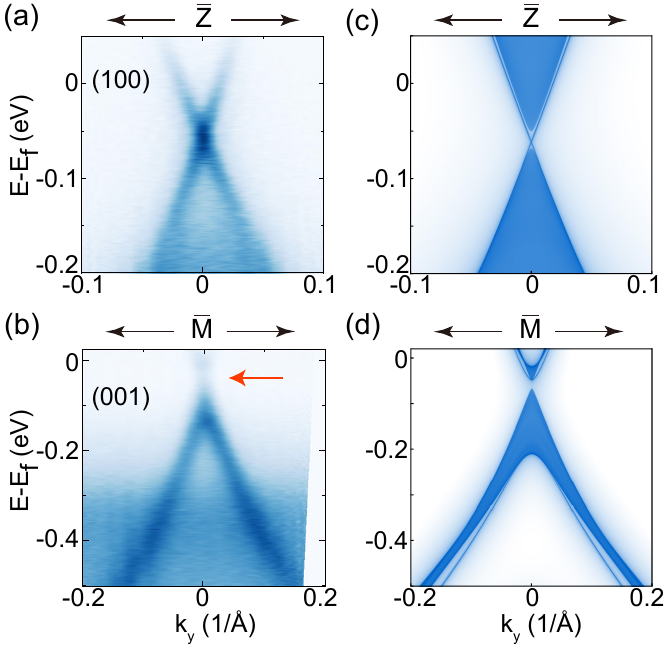}
\caption{Comparison between ARPES spectra and calculated surface states of $\beta$-Bi$_4$I$_4$ at 300~K. ARPES spectra around (a) the $\bar{Z}$ point on the (100) surface and (b) the $\bar{M}$ point on the (001) surface, taken from Ref.~\cite{Zhao2024}. The corresponding calculated surface states on (c) the (100) and (d) the (001) surface.}
\label{Fig:fig3}
\end{figure}

\textit{Phonon-induced topological phase transition}.---We incorporate phonon effects into our electronic structure calculations through e-ph self-energy corrections. 
Figure~\ref{Fig:fig2}(a) shows the phonon-renormalized band structures of $\beta$-Bi$_4$I$_4$ at various temperatures. 
Near the $M$ point, the conduction band shifts upward with increasing temperature, while the valence band remains nearly unchanged.
As a result, the local gap increases with temperature, opposite to the temperature-induced gap narrowing commonly observed in semiconductors.
At the $L$ point,  the conduction band minimum (CBM) and valence band maximum (VBM) initially move toward each other, leading to an almost closed gap (< 1~meV) at 180~K. The gap reopens upon further heating to 300~K.
Figure~\ref{Fig:fig2}(b) quantifies the temperature dependence of the band gap at the $L$ point.
Even at $T$ = 0~K, zero-point renormalization reduces the gap from the DFT value of 30~meV to $\sim$24~meV. 
As temperature rises, thermal phonons further suppress the gap until closure near 180~K, the gap then reopens and grows to $\sim$36~meV at 400~K.
This sequence of gap closure and reopening marks a critical point where band inversion and associated topological phase transition occur.\\
%
\indent
To confirm the band inversion, we examine the parity and orbital-projected band character of band-edge states across the critical point. 
At 100~K [Fig.~\ref{Fig:fig2}(c)], the VBM at the $L$ point with odd parity is dominated by Bi$_\text{in}$-$p_x$ character, while the CBM with even parity exhibits mainly Bi$_\text{ex}$-$p_x$ character. 
By 300~K [Fig.~\ref{Fig:fig2}(d)], both the parities and the orbital characters are reversed, providing definitive evidence of band inversion. 
As a result, the Z$_2$ invariants change from (1;110) to (0;001), confirming a topological phase transition from a strong to a weak TI at a critical temperature of $\approx$180~K [Fig.~\ref{Fig:fig2}(b)]. \\
\indent
Importantly, the topological transition temperature lies well below the structural transition temperature of $\sim$300~K. This separation of transition temperatures ensures that $\beta$-Bi$_4$I$_4$ resides exclusively in the weak TI regime under experimentally accessible conditions, as indicated by the shaded region in Fig.~\ref{Fig:fig2}(b). Notably, this topological phase transition is driven solely by thermal phonons, as all calculations are based on the experimental crystal structure, which exhibits negligible thermal expansion over the relevant temperature range~\cite{Huang2021}. This phonon-driven mechanism thus provides the crucial missing link between static-lattice predictions and experimental observations. \\
\indent
We further validate the room-temperature topological phase by comparing calculated surface states with ARPES spectra, as shown in Fig.~\ref{Fig:fig3}.
A key diagnostic for distinguishing the weak and strong TI phases in $\beta$-Bi$_4$I$_4$  is the presence (absence) of gapless TSS at the $\bar{Z}$-point ($\bar{M}$-point) on the (100) side [(001) top] surface~\cite{Autes2016,Noguchi2019}.
Our calculations predict gapless TSS at the $\bar{Z}$-point on the (100) surface, exhibiting Dirac-like linear dispersion across the bulk gap [Fig.~\ref{Fig:fig3}(c)]. The ARPES spectrum shown in Fig.~\ref{Fig:fig3}(a) reveals similar gapless states.
Note that these spectra were obtained using surface-sensitive nano-ARPES~\cite{Zhao2023,Zhao2024}, which primarily probes surface states and suppresses bulk contributions---explaining the near absence of bulk states features in Fig.~\ref{Fig:fig3}(a).
Equally important, the ARPES spectrum at the $\bar{M}$-point on the (001) surface shows clearly gapped states [Fig.~\ref{Fig:fig3}(b)], consistent with our calculation [Fig.~\ref{Fig:fig3}(d)], which reproduces spectral features of the gapped states with high fidelity.
Additional comparisons and the temperature dependence of the calculated surface states are provided in the SM~\cite{supmat}.
Taken together, these results provide unambiguous evidence that $\beta$-Bi$_4$I$_4$ at 300~K is a weak TI with Z$_2$ = (0;001), and confirm that e-ph renormalization is essential for correctly predicting the topological nature of $\beta$-Bi$_4$I$_4$.

\begin{figure}[!htbp]
\centering
\includegraphics[width=0.92\columnwidth]{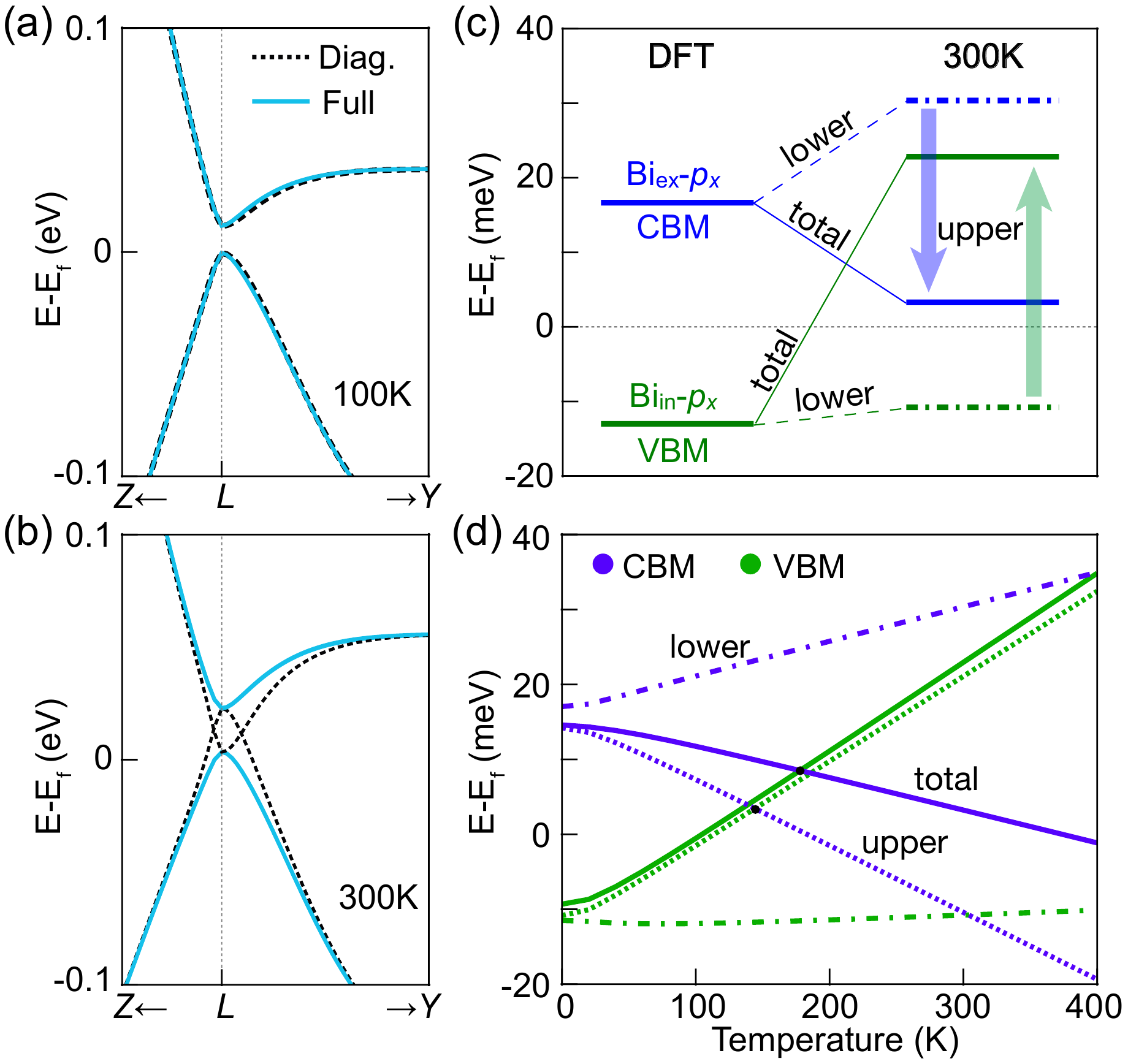}
\caption{Phonon-renormalized band structures computed with (solid) and without (dashed) off-diagonal self-energies at (a) $T = 100$~K and (b) $T = 300$~K. (c) Schematic illustration of the renormalized CBM and VBM energies at 300~K, computed with contributions from the lower-subspace only and the total. Arrows highlight the upper-subspace contribution that drives the inversion. (d) The renormalized CBM and VBM energies versus temperature, computed with contributions from only the lower-subspace (dash-dotted), upper-subspace (dotted), and total (solid).}
\label{Fig:fig4}
\end{figure}

Finally, we identify two critical mechanisms underlying the phonon-driven topological phase transition in $\beta$-Bi$_4$I$_4$.
The first one is phonon-induced wavefunction hybridization enabled by off-diagonal terms of the e-ph self-energy, which allows mixing of near-gap states~\cite{Lihm2020}. 
Figures~\ref{Fig:fig4}(a) and~\ref{Fig:fig4}(b) demonstrate the key role of these off-diagonal terms for reopenning the gap after band inversion. 
At 100~K (before inversion), band structures with and without these terms are nearly identical, indicating negligible hybridization. 
At 300~K (after inversion), however, these terms become indispensable: omitting them leaves gapless crossings and a metallic state, whereas including them hybridizes the inverted states, reopens the gap, and stabilizes the weak TI phase. 
Thus, these off-diagonal self-energies drive the gap-reopening process in this transition. \\
\indent
The second one is that the band inversion is driven entirely by phonon-mediated coupling between band edges and high-energy states far from the gap, namely the upper-subspace contribution to the e-ph self-energy, as illustrated in Fig.~\ref{Fig:fig4}(c). 
Like the Fan self-energy, the DW self-energy can be partitioned into lower- and upper-subspace contributions, with the lower subspace comprising 48 bands spanning from $-6$~eV to $+4$~eV of the gap (see Fig.~S1 in SM~\cite{supmat}).
We obtain the lower- and upper-subspace contributions to the total self-energy by summing the respective Fan and DW terms.
Figure~\ref{Fig:fig4}(d) shows the renormalized CBM and VBM energies as a function of temperature, using only the lower part, upper part, and total self-energy. 
The lower part produces minor shifts without inversion, whereas the upper part drives substantial shifts and band inversion near 140~K (compared to 180~K for the total).
This challenges the conventional wisdom that near-gap states dominate~\cite{Antonius2016}, and indicates high-energy states can strongly influence low-energy topology via phonon-mediated processes.

In summary, we resolved the long-standing discrepancy between theory and experiment for the topological phase of $\beta$-Bi$_4$I$_4$ by identifying e-ph renormalization as the crucial missing ingredient in prior theoretical models. 
Our results demonstrate that the room-temperature topological phase transition in Bi$_4$I$_4$ cannot be explained by structural changes alone, but fundamentally
requires temperature-activated phonons, which shape the low-energy topology by mediating coupling to high-energy states.
The \textit{ab initio} framework developed here enables accurate finite-temperature predictions, offering both fundamental insights and practical guidance for designing topological materials operating under realistic conditions.

\begin{acknowledgments} 
\textit{Acknowledgments}.---The authors acknowledge support from the National Key R\&D Program of China (Grant No.~2022YFA1403400), the National Natural Science Foundation of China (Grant Nos.~12104039, 12321004, 12234003), and the Beijing Natural Science Foundation (Grant No.~Z210006). 
\end{acknowledgments}

\bibliography{bii_refs}

\begin{thebibliography}{67}%
\makeatletter
\providecommand \@ifxundefined [1]{%
 \@ifx{#1\undefined}
}%
\providecommand \@ifnum [1]{%
 \ifnum #1\expandafter \@firstoftwo
 \else \expandafter \@secondoftwo
 \fi
}%
\providecommand \@ifx [1]{%
 \ifx #1\expandafter \@firstoftwo
 \else \expandafter \@secondoftwo
 \fi
}%
\providecommand \natexlab [1]{#1}%
\providecommand \enquote  [1]{``#1''}%
\providecommand \bibnamefont  [1]{#1}%
\providecommand \bibfnamefont [1]{#1}%
\providecommand \citenamefont [1]{#1}%
\providecommand \href@noop [0]{\@secondoftwo}%
\providecommand \href [0]{\begingroup \@sanitize@url \@href}%
\providecommand \@href[1]{\@@startlink{#1}\@@href}%
\providecommand \@@href[1]{\endgroup#1\@@endlink}%
\providecommand \@sanitize@url [0]{\catcode `\\12\catcode `\$12\catcode
  `\&12\catcode `\#12\catcode `\^12\catcode `\_12\catcode `\%12\relax}%
\providecommand \@@startlink[1]{}%
\providecommand \@@endlink[0]{}%
\providecommand \url  [0]{\begingroup\@sanitize@url \@url }%
\providecommand \@url [1]{\endgroup\@href {#1}{\urlprefix }}%
\providecommand \urlprefix  [0]{URL }%
\providecommand \Eprint [0]{\href }%
\providecommand \doibase [0]{https://doi.org/}%
\providecommand \selectlanguage [0]{\@gobble}%
\providecommand \bibinfo  [0]{\@secondoftwo}%
\providecommand \bibfield  [0]{\@secondoftwo}%
\providecommand \translation [1]{[#1]}%
\providecommand \BibitemOpen [0]{}%
\providecommand \bibitemStop [0]{}%
\providecommand \bibitemNoStop [0]{.\EOS\space}%
\providecommand \EOS [0]{\spacefactor3000\relax}%
\providecommand \BibitemShut  [1]{\csname bibitem#1\endcsname}%
\let\auto@bib@innerbib\@empty
\bibitem [{\citenamefont {Hasan}\ and\ \citenamefont {Kane}(2010)}]{Hasan2010}%
  \BibitemOpen
  \bibfield  {author} {\bibinfo {author} {\bibfnamefont {M.~Z.}\ \bibnamefont
  {Hasan}}\ and\ \bibinfo {author} {\bibfnamefont {C.~L.}\ \bibnamefont
  {Kane}},\ }\bibfield  {title} {\bibinfo {title} {Colloquium: Topological
  insulators},\ }\href {https://doi.org/10.1103/RevModPhys.82.3045} {\bibfield
  {journal} {\bibinfo  {journal} {Rev. Mod. Phys.}\ }\textbf {\bibinfo {volume}
  {82}},\ \bibinfo {pages} {3045} (\bibinfo {year} {2010})}\BibitemShut
  {NoStop}%
\bibitem [{\citenamefont {Qi}\ and\ \citenamefont {Zhang}(2011)}]{Qi2011}%
  \BibitemOpen
  \bibfield  {author} {\bibinfo {author} {\bibfnamefont {X.-L.}\ \bibnamefont
  {Qi}}\ and\ \bibinfo {author} {\bibfnamefont {S.-C.}\ \bibnamefont {Zhang}},\
  }\bibfield  {title} {\bibinfo {title} {Topological insulators and
  superconductors},\ }\href {https://doi.org/10.1103/RevModPhys.83.1057}
  {\bibfield  {journal} {\bibinfo  {journal} {Rev. Mod. Phys.}\ }\textbf
  {\bibinfo {volume} {83}},\ \bibinfo {pages} {1057} (\bibinfo {year}
  {2011})}\BibitemShut {NoStop}%
\bibitem [{\citenamefont {Bansil}\ \emph {et~al.}(2016)\citenamefont {Bansil},
  \citenamefont {Lin},\ and\ \citenamefont {Das}}]{Bansil2016}%
  \BibitemOpen
  \bibfield  {author} {\bibinfo {author} {\bibfnamefont {A.}~\bibnamefont
  {Bansil}}, \bibinfo {author} {\bibfnamefont {H.}~\bibnamefont {Lin}},\ and\
  \bibinfo {author} {\bibfnamefont {T.}~\bibnamefont {Das}},\ }\bibfield
  {title} {\bibinfo {title} {Colloquium: Topological band theory},\ }\href
  {https://doi.org/10.1103/RevModPhys.88.021004} {\bibfield  {journal}
  {\bibinfo  {journal} {Rev. Mod. Phys.}\ }\textbf {\bibinfo {volume} {88}},\
  \bibinfo {pages} {021004} (\bibinfo {year} {2016})}\BibitemShut {NoStop}%
\bibitem [{\citenamefont {Ren}\ \emph {et~al.}(2016)\citenamefont {Ren},
  \citenamefont {Qiao},\ and\ \citenamefont {Niu}}]{Ren2016}%
  \BibitemOpen
  \bibfield  {author} {\bibinfo {author} {\bibfnamefont {Y.}~\bibnamefont
  {Ren}}, \bibinfo {author} {\bibfnamefont {Z.}~\bibnamefont {Qiao}},\ and\
  \bibinfo {author} {\bibfnamefont {Q.}~\bibnamefont {Niu}},\ }\bibfield
  {title} {\bibinfo {title} {Topological phases in two-dimensional materials: a
  review},\ }\href {https://doi.org/10.1088/0034-4885/79/6/066501} {\bibfield
  {journal} {\bibinfo  {journal} {Rep. Progr. Phys.}\ }\textbf {\bibinfo
  {volume} {79}},\ \bibinfo {pages} {066501} (\bibinfo {year}
  {2016})}\BibitemShut {NoStop}%
\bibitem [{\citenamefont {Han}\ \emph {et~al.}(2022)\citenamefont {Han},
  \citenamefont {Xiao},\ and\ \citenamefont {Yao}}]{Han2022}%
  \BibitemOpen
  \bibfield  {author} {\bibinfo {author} {\bibfnamefont {J.}~\bibnamefont
  {Han}}, \bibinfo {author} {\bibfnamefont {W.}~\bibnamefont {Xiao}},\ and\
  \bibinfo {author} {\bibfnamefont {Y.}~\bibnamefont {Yao}},\ }\bibfield
  {title} {\bibinfo {title} {Quasi-one-dimensional topological material
  {Bi$_4$X$_4$}({X}={Br},{I})},\ }\href
  {https://doi.org/10.1080/23746149.2022.2057234} {\bibfield  {journal}
  {\bibinfo  {journal} {Adv. Phys.:X}\ }\textbf {\bibinfo {volume} {7}},\
  \bibinfo {pages} {2057234} (\bibinfo {year} {2022})}\BibitemShut {NoStop}%
\bibitem [{\citenamefont {Zhou}\ \emph {et~al.}(2014)\citenamefont {Zhou},
  \citenamefont {Feng}, \citenamefont {Liu}, \citenamefont {Guan},\ and\
  \citenamefont {Yao}}]{Zhou2014}%
  \BibitemOpen
  \bibfield  {author} {\bibinfo {author} {\bibfnamefont {J.-J.}\ \bibnamefont
  {Zhou}}, \bibinfo {author} {\bibfnamefont {W.}~\bibnamefont {Feng}}, \bibinfo
  {author} {\bibfnamefont {C.-C.}\ \bibnamefont {Liu}}, \bibinfo {author}
  {\bibfnamefont {S.}~\bibnamefont {Guan}},\ and\ \bibinfo {author}
  {\bibfnamefont {Y.}~\bibnamefont {Yao}},\ }\bibfield  {title} {\bibinfo
  {title} {Large-{Gap} {Quantum} {Spin} {Hall} {Insulator} in {Single} {Layer}
  {Bismuth} {Monobromide} {Bi$_4$Br$_4$}},\ }\href
  {https://doi.org/10.1021/nl501907g} {\bibfield  {journal} {\bibinfo
  {journal} {Nano Lett.}\ }\textbf {\bibinfo {volume} {14}},\ \bibinfo {pages}
  {4767} (\bibinfo {year} {2014})}\BibitemShut {NoStop}%
\bibitem [{\citenamefont {Zhou}\ \emph {et~al.}(2015)\citenamefont {Zhou},
  \citenamefont {Feng}, \citenamefont {Liu},\ and\ \citenamefont
  {Yao}}]{Zhou2015}%
  \BibitemOpen
  \bibfield  {author} {\bibinfo {author} {\bibfnamefont {J.-J.}\ \bibnamefont
  {Zhou}}, \bibinfo {author} {\bibfnamefont {W.}~\bibnamefont {Feng}}, \bibinfo
  {author} {\bibfnamefont {G.-B.}\ \bibnamefont {Liu}},\ and\ \bibinfo {author}
  {\bibfnamefont {Y.}~\bibnamefont {Yao}},\ }\bibfield  {title} {\bibinfo
  {title} {Topological edge states in single- and multi-layer {Bi$_4$Br$_4$}},\
  }\href {https://doi.org/10.1088/1367-2630/17/1/015004} {\bibfield  {journal}
  {\bibinfo  {journal} {New J. Phys.}\ }\textbf {\bibinfo {volume} {17}},\
  \bibinfo {pages} {015004} (\bibinfo {year} {2015})}\BibitemShut {NoStop}%
\bibitem [{\citenamefont {Han}\ \emph {et~al.}(2023)\citenamefont {Han},
  \citenamefont {Mao}, \citenamefont {Chen}, \citenamefont {Yin}, \citenamefont
  {Wang}, \citenamefont {Chen}, \citenamefont {Li}, \citenamefont {Zheng},
  \citenamefont {Zhang}, \citenamefont {Ma} \emph {et~al.}}]{Han2023}%
  \BibitemOpen
  \bibfield  {author} {\bibinfo {author} {\bibfnamefont {J.}~\bibnamefont
  {Han}}, \bibinfo {author} {\bibfnamefont {P.}~\bibnamefont {Mao}}, \bibinfo
  {author} {\bibfnamefont {H.}~\bibnamefont {Chen}}, \bibinfo {author}
  {\bibfnamefont {J.-X.}\ \bibnamefont {Yin}}, \bibinfo {author} {\bibfnamefont
  {M.}~\bibnamefont {Wang}}, \bibinfo {author} {\bibfnamefont {D.}~\bibnamefont
  {Chen}}, \bibinfo {author} {\bibfnamefont {Y.}~\bibnamefont {Li}}, \bibinfo
  {author} {\bibfnamefont {J.}~\bibnamefont {Zheng}}, \bibinfo {author}
  {\bibfnamefont {X.}~\bibnamefont {Zhang}}, \bibinfo {author} {\bibfnamefont
  {D.}~\bibnamefont {Ma}}, \emph {et~al.},\ }\bibfield  {title} {\bibinfo
  {title} {Optical bulk-boundary dichotomy in a quantum spin {Hall}
  insulator},\ }\href
  {https://doi.org/https://doi.org/10.1016/j.scib.2023.01.038} {\bibfield
  {journal} {\bibinfo  {journal} {Sci. Bull.}\ }\textbf {\bibinfo {volume}
  {68}},\ \bibinfo {pages} {417} (\bibinfo {year} {2023})}\BibitemShut
  {NoStop}%
\bibitem [{\citenamefont {Shumiya}\ \emph {et~al.}(2022)\citenamefont
  {Shumiya}, \citenamefont {Hossain}, \citenamefont {Yin}, \citenamefont
  {Wang}, \citenamefont {Litskevich}, \citenamefont {Yoon}, \citenamefont {Li},
  \citenamefont {Yang}, \citenamefont {Jiang}, \citenamefont {Cheng} \emph
  {et~al.}}]{Shumiya2022}%
  \BibitemOpen
  \bibfield  {author} {\bibinfo {author} {\bibfnamefont {N.}~\bibnamefont
  {Shumiya}}, \bibinfo {author} {\bibfnamefont {M.~S.}\ \bibnamefont
  {Hossain}}, \bibinfo {author} {\bibfnamefont {J.-X.}\ \bibnamefont {Yin}},
  \bibinfo {author} {\bibfnamefont {Z.}~\bibnamefont {Wang}}, \bibinfo {author}
  {\bibfnamefont {M.}~\bibnamefont {Litskevich}}, \bibinfo {author}
  {\bibfnamefont {C.}~\bibnamefont {Yoon}}, \bibinfo {author} {\bibfnamefont
  {Y.}~\bibnamefont {Li}}, \bibinfo {author} {\bibfnamefont {Y.}~\bibnamefont
  {Yang}}, \bibinfo {author} {\bibfnamefont {Y.-X.}\ \bibnamefont {Jiang}},
  \bibinfo {author} {\bibfnamefont {G.}~\bibnamefont {Cheng}}, \emph {et~al.},\
  }\bibfield  {title} {\bibinfo {title} {Evidence of a room-temperature quantum
  spin {Hall} edge state in a higher-order topological insulator},\ }\href
  {https://doi.org/10.1038/s41563-022-01304-3} {\bibfield  {journal} {\bibinfo
  {journal} {Nat. Mater.}\ }\textbf {\bibinfo {volume} {21}},\ \bibinfo {pages}
  {1111} (\bibinfo {year} {2022})}\BibitemShut {NoStop}%
\bibitem [{\citenamefont {Yang}\ \emph {et~al.}(2022)\citenamefont {Yang},
  \citenamefont {Liu}, \citenamefont {Zhou}, \citenamefont {Liu}, \citenamefont
  {Mu}, \citenamefont {Liu}, \citenamefont {Wang}, \citenamefont {Hao},
  \citenamefont {Li}, \citenamefont {Zhong} \emph {et~al.}}]{Yang2022}%
  \BibitemOpen
  \bibfield  {author} {\bibinfo {author} {\bibfnamefont {M.}~\bibnamefont
  {Yang}}, \bibinfo {author} {\bibfnamefont {Y.}~\bibnamefont {Liu}}, \bibinfo
  {author} {\bibfnamefont {W.}~\bibnamefont {Zhou}}, \bibinfo {author}
  {\bibfnamefont {C.}~\bibnamefont {Liu}}, \bibinfo {author} {\bibfnamefont
  {D.}~\bibnamefont {Mu}}, \bibinfo {author} {\bibfnamefont {Y.}~\bibnamefont
  {Liu}}, \bibinfo {author} {\bibfnamefont {J.}~\bibnamefont {Wang}}, \bibinfo
  {author} {\bibfnamefont {W.}~\bibnamefont {Hao}}, \bibinfo {author}
  {\bibfnamefont {J.}~\bibnamefont {Li}}, \bibinfo {author} {\bibfnamefont
  {J.}~\bibnamefont {Zhong}}, \emph {et~al.},\ }\bibfield  {title} {\bibinfo
  {title} {Large-{Gap} {Quantum} {Spin} {Hall} {State} and
  {Temperature}-{Induced} {Lifshitz} {Transition} in {Bi$_4$Br$_4$}},\ }\href
  {https://doi.org/10.1021/acsnano.1c10539} {\bibfield  {journal} {\bibinfo
  {journal} {ACS Nano}\ }\textbf {\bibinfo {volume} {16}},\ \bibinfo {pages}
  {3036} (\bibinfo {year} {2022})}\BibitemShut {NoStop}%
\bibitem [{\citenamefont {Peng}\ \emph {et~al.}(2021)\citenamefont {Peng},
  \citenamefont {Zhang}, \citenamefont {Dong}, \citenamefont {Ma},
  \citenamefont {Chen}, \citenamefont {Li}, \citenamefont {Li}, \citenamefont
  {Han}, \citenamefont {Wang}, \citenamefont {Liu}, \citenamefont {Zhou},
  \citenamefont {Xiao},\ and\ \citenamefont {Yao}}]{Peng2021}%
  \BibitemOpen
  \bibfield  {author} {\bibinfo {author} {\bibfnamefont {X.}~\bibnamefont
  {Peng}}, \bibinfo {author} {\bibfnamefont {X.}~\bibnamefont {Zhang}},
  \bibinfo {author} {\bibfnamefont {X.}~\bibnamefont {Dong}}, \bibinfo {author}
  {\bibfnamefont {D.}~\bibnamefont {Ma}}, \bibinfo {author} {\bibfnamefont
  {D.}~\bibnamefont {Chen}}, \bibinfo {author} {\bibfnamefont {Y.}~\bibnamefont
  {Li}}, \bibinfo {author} {\bibfnamefont {J.}~\bibnamefont {Li}}, \bibinfo
  {author} {\bibfnamefont {J.}~\bibnamefont {Han}}, \bibinfo {author}
  {\bibfnamefont {Z.}~\bibnamefont {Wang}}, \bibinfo {author} {\bibfnamefont
  {C.-C.}\ \bibnamefont {Liu}}, \bibinfo {author} {\bibfnamefont
  {J.}~\bibnamefont {Zhou}}, \bibinfo {author} {\bibfnamefont {W.}~\bibnamefont
  {Xiao}},\ and\ \bibinfo {author} {\bibfnamefont {Y.}~\bibnamefont {Yao}},\
  }\bibfield  {title} {\bibinfo {title} {Observation of {Topological} {Edge}
  {States} on $\alpha$-{Bi$_4$Br$_4$} {Nanowires} {Grown} on {TiSe$_2$}
  {Substrates}},\ }\href {https://doi.org/10.1021/acs.jpclett.1c02586}
  {\bibfield  {journal} {\bibinfo  {journal} {J. Phys. Chem. Lett.}\ }\textbf
  {\bibinfo {volume} {12}},\ \bibinfo {pages} {10465} (\bibinfo {year}
  {2021})}\BibitemShut {NoStop}%
\bibitem [{\citenamefont {Noguchi}\ \emph {et~al.}(2019)\citenamefont
  {Noguchi}, \citenamefont {Takahashi}, \citenamefont {Kuroda}, \citenamefont
  {Ochi}, \citenamefont {Shirasawa}, \citenamefont {Sakano}, \citenamefont
  {Bareille}, \citenamefont {Nakayama}, \citenamefont {Watson}, \citenamefont
  {Yaji} \emph {et~al.}}]{Noguchi2019}%
  \BibitemOpen
  \bibfield  {author} {\bibinfo {author} {\bibfnamefont {R.}~\bibnamefont
  {Noguchi}}, \bibinfo {author} {\bibfnamefont {T.}~\bibnamefont {Takahashi}},
  \bibinfo {author} {\bibfnamefont {K.}~\bibnamefont {Kuroda}}, \bibinfo
  {author} {\bibfnamefont {M.}~\bibnamefont {Ochi}}, \bibinfo {author}
  {\bibfnamefont {T.}~\bibnamefont {Shirasawa}}, \bibinfo {author}
  {\bibfnamefont {M.}~\bibnamefont {Sakano}}, \bibinfo {author} {\bibfnamefont
  {C.}~\bibnamefont {Bareille}}, \bibinfo {author} {\bibfnamefont
  {M.}~\bibnamefont {Nakayama}}, \bibinfo {author} {\bibfnamefont {M.~D.}\
  \bibnamefont {Watson}}, \bibinfo {author} {\bibfnamefont {K.}~\bibnamefont
  {Yaji}}, \emph {et~al.},\ }\bibfield  {title} {\bibinfo {title} {A weak
  topological insulator state in quasi-one-dimensional bismuth iodide},\ }\href
  {https://doi.org/10.1038/s41586-019-0927-7} {\bibfield  {journal} {\bibinfo
  {journal} {Nature}\ }\textbf {\bibinfo {volume} {566}},\ \bibinfo {pages}
  {518} (\bibinfo {year} {2019})}\BibitemShut {NoStop}%
\bibitem [{\citenamefont {Huang}\ \emph {et~al.}(2021)\citenamefont {Huang},
  \citenamefont {Li}, \citenamefont {Yoon}, \citenamefont {Oh}, \citenamefont
  {Wu}, \citenamefont {Liu}, \citenamefont {Dhale}, \citenamefont {Zhou},
  \citenamefont {Guo}, \citenamefont {Zhang} \emph {et~al.}}]{Huang2021}%
  \BibitemOpen
  \bibfield  {author} {\bibinfo {author} {\bibfnamefont {J.}~\bibnamefont
  {Huang}}, \bibinfo {author} {\bibfnamefont {S.}~\bibnamefont {Li}}, \bibinfo
  {author} {\bibfnamefont {C.}~\bibnamefont {Yoon}}, \bibinfo {author}
  {\bibfnamefont {J.~S.}\ \bibnamefont {Oh}}, \bibinfo {author} {\bibfnamefont
  {H.}~\bibnamefont {Wu}}, \bibinfo {author} {\bibfnamefont {X.}~\bibnamefont
  {Liu}}, \bibinfo {author} {\bibfnamefont {N.}~\bibnamefont {Dhale}}, \bibinfo
  {author} {\bibfnamefont {Y.-F.}\ \bibnamefont {Zhou}}, \bibinfo {author}
  {\bibfnamefont {Y.}~\bibnamefont {Guo}}, \bibinfo {author} {\bibfnamefont
  {Y.}~\bibnamefont {Zhang}}, \emph {et~al.},\ }\bibfield  {title} {\bibinfo
  {title} {Room-{Temperature} {Topological} {Phase} {Transition} in
  {Quasi}-{One}-{Dimensional} {Material} {Bi$_4$I$_4$}},\ }\href
  {https://doi.org/10.1103/PhysRevX.11.031042} {\bibfield  {journal} {\bibinfo
  {journal} {Phys. Rev. X}\ }\textbf {\bibinfo {volume} {11}},\ \bibinfo
  {pages} {031042} (\bibinfo {year} {2021})}\BibitemShut {NoStop}%
\bibitem [{\citenamefont {Zhao}\ \emph {et~al.}(2024)\citenamefont {Zhao},
  \citenamefont {Yang}, \citenamefont {Du}, \citenamefont {Li}, \citenamefont
  {Zhai}, \citenamefont {Hu}, \citenamefont {Han}, \citenamefont {Huang},
  \citenamefont {Liu}, \citenamefont {Yao} \emph {et~al.}}]{Zhao2024}%
  \BibitemOpen
  \bibfield  {author} {\bibinfo {author} {\bibfnamefont {W.~X.}\ \bibnamefont
  {Zhao}}, \bibinfo {author} {\bibfnamefont {M.}~\bibnamefont {Yang}}, \bibinfo
  {author} {\bibfnamefont {X.}~\bibnamefont {Du}}, \bibinfo {author}
  {\bibfnamefont {Y.~D.}\ \bibnamefont {Li}}, \bibinfo {author} {\bibfnamefont
  {K.~Y.}\ \bibnamefont {Zhai}}, \bibinfo {author} {\bibfnamefont {Y.~Q.}\
  \bibnamefont {Hu}}, \bibinfo {author} {\bibfnamefont {J.~F.}\ \bibnamefont
  {Han}}, \bibinfo {author} {\bibfnamefont {Y.}~\bibnamefont {Huang}}, \bibinfo
  {author} {\bibfnamefont {Z.~K.}\ \bibnamefont {Liu}}, \bibinfo {author}
  {\bibfnamefont {Y.~G.}\ \bibnamefont {Yao}}, \emph {et~al.},\ }\bibfield
  {title} {\bibinfo {title} {Topological phase transition in
  quasi-one-dimensional bismuth iodide {Bi$_4$I$_4$}},\ }\href
  {https://doi.org/10.1038/s41535-024-00711-w} {\bibfield  {journal} {\bibinfo
  {journal} {npj Quantum Mater.}\ }\textbf {\bibinfo {volume} {9}},\ \bibinfo
  {pages} {1} (\bibinfo {year} {2024})}\BibitemShut {NoStop}%
\bibitem [{\citenamefont {Noguchi}\ \emph {et~al.}(2021)\citenamefont
  {Noguchi}, \citenamefont {Kobayashi}, \citenamefont {Jiang}, \citenamefont
  {Kuroda}, \citenamefont {Takahashi}, \citenamefont {Xu}, \citenamefont {Lee},
  \citenamefont {Hirayama}, \citenamefont {Ochi}, \citenamefont {Shirasawa}
  \emph {et~al.}}]{Noguchi2021}%
  \BibitemOpen
  \bibfield  {author} {\bibinfo {author} {\bibfnamefont {R.}~\bibnamefont
  {Noguchi}}, \bibinfo {author} {\bibfnamefont {M.}~\bibnamefont {Kobayashi}},
  \bibinfo {author} {\bibfnamefont {Z.}~\bibnamefont {Jiang}}, \bibinfo
  {author} {\bibfnamefont {K.}~\bibnamefont {Kuroda}}, \bibinfo {author}
  {\bibfnamefont {T.}~\bibnamefont {Takahashi}}, \bibinfo {author}
  {\bibfnamefont {Z.}~\bibnamefont {Xu}}, \bibinfo {author} {\bibfnamefont
  {D.}~\bibnamefont {Lee}}, \bibinfo {author} {\bibfnamefont {M.}~\bibnamefont
  {Hirayama}}, \bibinfo {author} {\bibfnamefont {M.}~\bibnamefont {Ochi}},
  \bibinfo {author} {\bibfnamefont {T.}~\bibnamefont {Shirasawa}}, \emph
  {et~al.},\ }\bibfield  {title} {\bibinfo {title} {Evidence for a higher-order
  topological insulator in a three-dimensional material built from van der
  {Waals} stacking of bismuth-halide chains},\ }\href
  {https://doi.org/10.1038/s41563-020-00871-7} {\bibfield  {journal} {\bibinfo
  {journal} {Nat. Mater.}\ }\textbf {\bibinfo {volume} {20}},\ \bibinfo {pages}
  {473} (\bibinfo {year} {2021})}\BibitemShut {NoStop}%
\bibitem [{\citenamefont {Zhao}\ \emph {et~al.}(2023)\citenamefont {Zhao},
  \citenamefont {Yang}, \citenamefont {Xu}, \citenamefont {Du}, \citenamefont
  {Li}, \citenamefont {Zhai}, \citenamefont {Peng}, \citenamefont {Pei},
  \citenamefont {Gao}, \citenamefont {Li} \emph {et~al.}}]{Zhao2023}%
  \BibitemOpen
  \bibfield  {author} {\bibinfo {author} {\bibfnamefont {W.}~\bibnamefont
  {Zhao}}, \bibinfo {author} {\bibfnamefont {M.}~\bibnamefont {Yang}}, \bibinfo
  {author} {\bibfnamefont {R.}~\bibnamefont {Xu}}, \bibinfo {author}
  {\bibfnamefont {X.}~\bibnamefont {Du}}, \bibinfo {author} {\bibfnamefont
  {Y.}~\bibnamefont {Li}}, \bibinfo {author} {\bibfnamefont {K.}~\bibnamefont
  {Zhai}}, \bibinfo {author} {\bibfnamefont {C.}~\bibnamefont {Peng}}, \bibinfo
  {author} {\bibfnamefont {D.}~\bibnamefont {Pei}}, \bibinfo {author}
  {\bibfnamefont {H.}~\bibnamefont {Gao}}, \bibinfo {author} {\bibfnamefont
  {Y.}~\bibnamefont {Li}}, \emph {et~al.},\ }\bibfield  {title} {\bibinfo
  {title} {Topological electronic structure and spin texture of
  quasi-one-dimensional higher-order topological insulator {Bi$_4$Br$_4$}},\
  }\href {https://doi.org/10.1038/s41467-023-43882-z} {\bibfield  {journal}
  {\bibinfo  {journal} {Nat. Commun.}\ }\textbf {\bibinfo {volume} {14}},\
  \bibinfo {pages} {8089} (\bibinfo {year} {2023})}\BibitemShut {NoStop}%
\bibitem [{\citenamefont {Hossain}\ \emph {et~al.}(2024)\citenamefont
  {Hossain}, \citenamefont {Zhang}, \citenamefont {Wang}, \citenamefont
  {Dhale}, \citenamefont {Liu}, \citenamefont {Litskevich}, \citenamefont
  {Casas}, \citenamefont {Shumiya}, \citenamefont {Yin}, \citenamefont
  {Cochran} \emph {et~al.}}]{Hossain2024}%
  \BibitemOpen
  \bibfield  {author} {\bibinfo {author} {\bibfnamefont {M.~S.}\ \bibnamefont
  {Hossain}}, \bibinfo {author} {\bibfnamefont {Q.}~\bibnamefont {Zhang}},
  \bibinfo {author} {\bibfnamefont {Z.}~\bibnamefont {Wang}}, \bibinfo {author}
  {\bibfnamefont {N.}~\bibnamefont {Dhale}}, \bibinfo {author} {\bibfnamefont
  {W.}~\bibnamefont {Liu}}, \bibinfo {author} {\bibfnamefont {M.}~\bibnamefont
  {Litskevich}}, \bibinfo {author} {\bibfnamefont {B.}~\bibnamefont {Casas}},
  \bibinfo {author} {\bibfnamefont {N.}~\bibnamefont {Shumiya}}, \bibinfo
  {author} {\bibfnamefont {J.-X.}\ \bibnamefont {Yin}}, \bibinfo {author}
  {\bibfnamefont {T.~A.}\ \bibnamefont {Cochran}}, \emph {et~al.},\ }\bibfield
  {title} {\bibinfo {title} {Quantum transport response of topological hinge
  modes},\ }\href {https://doi.org/10.1038/s41567-024-02388-1} {\bibfield
  {journal} {\bibinfo  {journal} {Nat. Phys.}\ }\textbf {\bibinfo {volume}
  {20}},\ \bibinfo {pages} {776} (\bibinfo {year} {2024})}\BibitemShut
  {NoStop}%
\bibitem [{\citenamefont {Lefeuvre}\ \emph {et~al.}(2025)\citenamefont
  {Lefeuvre}, \citenamefont {Kobayashi}, \citenamefont {Patriarche},
  \citenamefont {Findling}, \citenamefont {Troadec}, \citenamefont {Ferrier},
  \citenamefont {Guéron}, \citenamefont {Bouchiat}, \citenamefont {Sasagawa},\
  and\ \citenamefont {Deblock}}]{Lefeuvre2025}%
  \BibitemOpen
  \bibfield  {author} {\bibinfo {author} {\bibfnamefont {J.}~\bibnamefont
  {Lefeuvre}}, \bibinfo {author} {\bibfnamefont {M.}~\bibnamefont {Kobayashi}},
  \bibinfo {author} {\bibfnamefont {G.}~\bibnamefont {Patriarche}}, \bibinfo
  {author} {\bibfnamefont {N.}~\bibnamefont {Findling}}, \bibinfo {author}
  {\bibfnamefont {D.}~\bibnamefont {Troadec}}, \bibinfo {author} {\bibfnamefont
  {M.}~\bibnamefont {Ferrier}}, \bibinfo {author} {\bibfnamefont
  {S.}~\bibnamefont {Guéron}}, \bibinfo {author} {\bibfnamefont
  {H.}~\bibnamefont {Bouchiat}}, \bibinfo {author} {\bibfnamefont
  {T.}~\bibnamefont {Sasagawa}},\ and\ \bibinfo {author} {\bibfnamefont
  {R.}~\bibnamefont {Deblock}},\ }\href
  {https://doi.org/10.48550/arXiv.2502.13837} {\bibinfo {title} {Quantum
  {Coherent} {Transport} of {1D} ballistic states in second order topological
  insulator {Bi$_4$Br$_4$}}} (\bibinfo {year} {2025}),\ \bibinfo {note}
  {\href{http://arxiv.org/abs/2502.13837}{arXiv:2502.13837
  [cond-mat]}}\BibitemShut {NoStop}%
\bibitem [{\citenamefont {Yu}\ \emph {et~al.}(2024)\citenamefont {Yu},
  \citenamefont {Deng}, \citenamefont {Liu}, \citenamefont {Zhang},
  \citenamefont {Sun}, \citenamefont {Dhale}, \citenamefont {Li}, \citenamefont
  {Ma}, \citenamefont {Wang}, \citenamefont {Wu} \emph {et~al.}}]{Yu2024}%
  \BibitemOpen
  \bibfield  {author} {\bibinfo {author} {\bibfnamefont {S.}~\bibnamefont
  {Yu}}, \bibinfo {author} {\bibfnamefont {J.}~\bibnamefont {Deng}}, \bibinfo
  {author} {\bibfnamefont {W.}~\bibnamefont {Liu}}, \bibinfo {author}
  {\bibfnamefont {Y.}~\bibnamefont {Zhang}}, \bibinfo {author} {\bibfnamefont
  {Y.}~\bibnamefont {Sun}}, \bibinfo {author} {\bibfnamefont {N.}~\bibnamefont
  {Dhale}}, \bibinfo {author} {\bibfnamefont {S.}~\bibnamefont {Li}}, \bibinfo
  {author} {\bibfnamefont {W.}~\bibnamefont {Ma}}, \bibinfo {author}
  {\bibfnamefont {Z.}~\bibnamefont {Wang}}, \bibinfo {author} {\bibfnamefont
  {P.}~\bibnamefont {Wu}}, \emph {et~al.},\ }\bibfield  {title} {\bibinfo
  {title} {Observation of {Robust} {One}-{Dimensional} {Edge} {Channels} in a
  {Three}-{Dimensional} {Quantum} {Spin} {Hall} {Insulator}},\ }\href
  {https://doi.org/10.1103/PhysRevX.14.041048} {\bibfield  {journal} {\bibinfo
  {journal} {Phys. Rev. X}\ }\textbf {\bibinfo {volume} {14}},\ \bibinfo
  {pages} {041048} (\bibinfo {year} {2024})}\BibitemShut {NoStop}%
\bibitem [{\citenamefont {Yang}\ \emph {et~al.}(2024)\citenamefont {Yang},
  \citenamefont {Zhao}, \citenamefont {Mu}, \citenamefont {Shi}, \citenamefont
  {Zhong}, \citenamefont {Li}, \citenamefont {Liu}, \citenamefont {Zhong},
  \citenamefont {Cheng}, \citenamefont {Zhou} \emph {et~al.}}]{Yang2024}%
  \BibitemOpen
  \bibfield  {author} {\bibinfo {author} {\bibfnamefont {M.}~\bibnamefont
  {Yang}}, \bibinfo {author} {\bibfnamefont {W.}~\bibnamefont {Zhao}}, \bibinfo
  {author} {\bibfnamefont {D.}~\bibnamefont {Mu}}, \bibinfo {author}
  {\bibfnamefont {Z.}~\bibnamefont {Shi}}, \bibinfo {author} {\bibfnamefont
  {J.}~\bibnamefont {Zhong}}, \bibinfo {author} {\bibfnamefont
  {Y.}~\bibnamefont {Li}}, \bibinfo {author} {\bibfnamefont {Y.}~\bibnamefont
  {Liu}}, \bibinfo {author} {\bibfnamefont {J.}~\bibnamefont {Zhong}}, \bibinfo
  {author} {\bibfnamefont {N.}~\bibnamefont {Cheng}}, \bibinfo {author}
  {\bibfnamefont {W.}~\bibnamefont {Zhou}}, \emph {et~al.},\ }\bibfield
  {title} {\bibinfo {title} {Mass {Acquisition} of {Dirac} {Fermions} in
  {Bi$_4$I$_4$} by {Spontaneous} {Symmetry} {Breaking}},\ }\href
  {https://doi.org/10.1103/PhysRevLett.133.256601} {\bibfield  {journal}
  {\bibinfo  {journal} {Phys. Rev. Lett.}\ }\textbf {\bibinfo {volume} {133}},\
  \bibinfo {pages} {256601} (\bibinfo {year} {2024})}\BibitemShut {NoStop}%
\bibitem [{\citenamefont {Zhong}\ \emph {et~al.}(2023)\citenamefont {Zhong},
  \citenamefont {Yang}, \citenamefont {Shi}, \citenamefont {Li}, \citenamefont
  {Mu}, \citenamefont {Liu}, \citenamefont {Cheng}, \citenamefont {Zhao},
  \citenamefont {Hao}, \citenamefont {Wang} \emph {et~al.}}]{Zhong2023}%
  \BibitemOpen
  \bibfield  {author} {\bibinfo {author} {\bibfnamefont {J.}~\bibnamefont
  {Zhong}}, \bibinfo {author} {\bibfnamefont {M.}~\bibnamefont {Yang}},
  \bibinfo {author} {\bibfnamefont {Z.}~\bibnamefont {Shi}}, \bibinfo {author}
  {\bibfnamefont {Y.}~\bibnamefont {Li}}, \bibinfo {author} {\bibfnamefont
  {D.}~\bibnamefont {Mu}}, \bibinfo {author} {\bibfnamefont {Y.}~\bibnamefont
  {Liu}}, \bibinfo {author} {\bibfnamefont {N.}~\bibnamefont {Cheng}}, \bibinfo
  {author} {\bibfnamefont {W.}~\bibnamefont {Zhao}}, \bibinfo {author}
  {\bibfnamefont {W.}~\bibnamefont {Hao}}, \bibinfo {author} {\bibfnamefont
  {J.}~\bibnamefont {Wang}}, \emph {et~al.},\ }\bibfield  {title} {\bibinfo
  {title} {Towards layer-selective quantum spin hall channels in weak
  topological insulator {Bi$_4$Br$_2$I$_2$}},\ }\href
  {https://doi.org/10.1038/s41467-023-40735-7} {\bibfield  {journal} {\bibinfo
  {journal} {Nat. Commun.}\ }\textbf {\bibinfo {volume} {14}},\ \bibinfo
  {pages} {4964} (\bibinfo {year} {2023})}\BibitemShut {NoStop}%
\bibitem [{\citenamefont {Noguchi}\ \emph {et~al.}(2024)\citenamefont
  {Noguchi}, \citenamefont {Kobayashi}, \citenamefont {Kawaguchi},
  \citenamefont {Yamamori}, \citenamefont {Aido}, \citenamefont {Lin},
  \citenamefont {Tanaka}, \citenamefont {Kuroda}, \citenamefont {Harasawa},
  \citenamefont {Kandyba} \emph {et~al.}}]{Noguchi2024}%
  \BibitemOpen
  \bibfield  {author} {\bibinfo {author} {\bibfnamefont {R.}~\bibnamefont
  {Noguchi}}, \bibinfo {author} {\bibfnamefont {M.}~\bibnamefont {Kobayashi}},
  \bibinfo {author} {\bibfnamefont {K.}~\bibnamefont {Kawaguchi}}, \bibinfo
  {author} {\bibfnamefont {W.}~\bibnamefont {Yamamori}}, \bibinfo {author}
  {\bibfnamefont {K.}~\bibnamefont {Aido}}, \bibinfo {author} {\bibfnamefont
  {C.}~\bibnamefont {Lin}}, \bibinfo {author} {\bibfnamefont {H.}~\bibnamefont
  {Tanaka}}, \bibinfo {author} {\bibfnamefont {K.}~\bibnamefont {Kuroda}},
  \bibinfo {author} {\bibfnamefont {A.}~\bibnamefont {Harasawa}}, \bibinfo
  {author} {\bibfnamefont {V.}~\bibnamefont {Kandyba}}, \emph {et~al.},\
  }\bibfield  {title} {\bibinfo {title} {Robust {Weak} {Topological}
  {Insulator} in the {Bismuth} {Halide} {Bi$_4$Br$_2$I$_2$}},\ }\href
  {https://doi.org/10.1103/PhysRevLett.133.086602} {\bibfield  {journal}
  {\bibinfo  {journal} {Phys. Rev. Lett.}\ }\textbf {\bibinfo {volume} {133}},\
  \bibinfo {pages} {086602} (\bibinfo {year} {2024})}\BibitemShut {NoStop}%
\bibitem [{\citenamefont {Zhong}\ \emph {et~al.}(2025)\citenamefont {Zhong},
  \citenamefont {Yang}, \citenamefont {Zhao}, \citenamefont {Zhai},
  \citenamefont {Zhen}, \citenamefont {Zhang}, \citenamefont {Mu},
  \citenamefont {Liu}, \citenamefont {Shi}, \citenamefont {Cheng} \emph
  {et~al.}}]{Zhong2025}%
  \BibitemOpen
  \bibfield  {author} {\bibinfo {author} {\bibfnamefont {J.}~\bibnamefont
  {Zhong}}, \bibinfo {author} {\bibfnamefont {M.}~\bibnamefont {Yang}},
  \bibinfo {author} {\bibfnamefont {W.}~\bibnamefont {Zhao}}, \bibinfo {author}
  {\bibfnamefont {K.}~\bibnamefont {Zhai}}, \bibinfo {author} {\bibfnamefont
  {X.}~\bibnamefont {Zhen}}, \bibinfo {author} {\bibfnamefont {L.}~\bibnamefont
  {Zhang}}, \bibinfo {author} {\bibfnamefont {D.}~\bibnamefont {Mu}}, \bibinfo
  {author} {\bibfnamefont {Y.}~\bibnamefont {Liu}}, \bibinfo {author}
  {\bibfnamefont {Z.}~\bibnamefont {Shi}}, \bibinfo {author} {\bibfnamefont
  {N.}~\bibnamefont {Cheng}}, \emph {et~al.},\ }\bibfield  {title} {\bibinfo
  {title} {Coalescence of multiple topological orders in quasi-one-dimensional
  bismuth halide chains},\ }\href {https://doi.org/10.1038/s41467-025-56593-4}
  {\bibfield  {journal} {\bibinfo  {journal} {Nat. Commun.}\ }\textbf {\bibinfo
  {volume} {16}},\ \bibinfo {pages} {1163} (\bibinfo {year}
  {2025})}\BibitemShut {NoStop}%
\bibitem [{\citenamefont {Hinostroza}\ \emph {et~al.}(2024)\citenamefont
  {Hinostroza}, \citenamefont {Rodrigues~de Faria}, \citenamefont {Cassemiro},
  \citenamefont {Larrea~Jiménez}, \citenamefont {Jefferson~da Silva~Machado},
  \citenamefont {Brito},\ and\ \citenamefont {Martelli}}]{Hinostroza2024}%
  \BibitemOpen
  \bibfield  {author} {\bibinfo {author} {\bibfnamefont {C.~D.}\ \bibnamefont
  {Hinostroza}}, \bibinfo {author} {\bibfnamefont {L.}~\bibnamefont
  {Rodrigues~de Faria}}, \bibinfo {author} {\bibfnamefont {G.~H.}\ \bibnamefont
  {Cassemiro}}, \bibinfo {author} {\bibfnamefont {J.}~\bibnamefont
  {Larrea~Jiménez}}, \bibinfo {author} {\bibfnamefont {A.}~\bibnamefont
  {Jefferson~da Silva~Machado}}, \bibinfo {author} {\bibfnamefont {W.~H.}\
  \bibnamefont {Brito}},\ and\ \bibinfo {author} {\bibfnamefont
  {V.}~\bibnamefont {Martelli}},\ }\bibfield  {title} {\bibinfo {title}
  {Structural investigation of the quasi-one-dimensional topological insulator
  {Bi$_4$I$_4$}},\ }\href {https://doi.org/10.1103/PhysRevB.109.174105}
  {\bibfield  {journal} {\bibinfo  {journal} {Phys. Rev. B}\ }\textbf {\bibinfo
  {volume} {109}},\ \bibinfo {pages} {174105} (\bibinfo {year}
  {2024})}\BibitemShut {NoStop}%
\bibitem [{\citenamefont {Liu}\ \emph {et~al.}(2016)\citenamefont {Liu},
  \citenamefont {Zhou}, \citenamefont {Yao},\ and\ \citenamefont
  {Zhang}}]{Liu2016}%
  \BibitemOpen
  \bibfield  {author} {\bibinfo {author} {\bibfnamefont {C.-C.}\ \bibnamefont
  {Liu}}, \bibinfo {author} {\bibfnamefont {J.-J.}\ \bibnamefont {Zhou}},
  \bibinfo {author} {\bibfnamefont {Y.}~\bibnamefont {Yao}},\ and\ \bibinfo
  {author} {\bibfnamefont {F.}~\bibnamefont {Zhang}},\ }\bibfield  {title}
  {\bibinfo {title} {Weak {Topological} {Insulators} and {Composite} {Weyl}
  {Semimetals}: {$\beta$-Bi$_4$X$_4$} {(X = Br, I)}},\ }\href
  {https://doi.org/10.1103/PhysRevLett.116.066801} {\bibfield  {journal}
  {\bibinfo  {journal} {Phys. Rev. Lett.}\ }\textbf {\bibinfo {volume} {116}},\
  \bibinfo {pages} {066801} (\bibinfo {year} {2016})}\BibitemShut {NoStop}%
\bibitem [{\citenamefont {Autès}\ \emph {et~al.}(2016)\citenamefont {Autès},
  \citenamefont {Isaeva}, \citenamefont {Moreschini}, \citenamefont
  {Johannsen}, \citenamefont {Pisoni}, \citenamefont {Mori}, \citenamefont
  {Zhang}, \citenamefont {Filatova}, \citenamefont {Kuznetsov}, \citenamefont
  {Forró} \emph {et~al.}}]{Autes2016}%
  \BibitemOpen
  \bibfield  {author} {\bibinfo {author} {\bibfnamefont {G.}~\bibnamefont
  {Autès}}, \bibinfo {author} {\bibfnamefont {A.}~\bibnamefont {Isaeva}},
  \bibinfo {author} {\bibfnamefont {L.}~\bibnamefont {Moreschini}}, \bibinfo
  {author} {\bibfnamefont {J.~C.}\ \bibnamefont {Johannsen}}, \bibinfo {author}
  {\bibfnamefont {A.}~\bibnamefont {Pisoni}}, \bibinfo {author} {\bibfnamefont
  {R.}~\bibnamefont {Mori}}, \bibinfo {author} {\bibfnamefont {W.}~\bibnamefont
  {Zhang}}, \bibinfo {author} {\bibfnamefont {T.~G.}\ \bibnamefont {Filatova}},
  \bibinfo {author} {\bibfnamefont {A.~N.}\ \bibnamefont {Kuznetsov}}, \bibinfo
  {author} {\bibfnamefont {L.}~\bibnamefont {Forró}}, \emph {et~al.},\
  }\bibfield  {title} {\bibinfo {title} {A novel quasi-one-dimensional
  topological insulator in bismuth iodide $\beta$-{Bi$_4$I$_4$}},\ }\href
  {https://doi.org/10.1038/nmat4488} {\bibfield  {journal} {\bibinfo  {journal}
  {Nat. Mater.}\ }\textbf {\bibinfo {volume} {15}},\ \bibinfo {pages} {154}
  (\bibinfo {year} {2016})}\BibitemShut {NoStop}%
\bibitem [{\citenamefont {Liu}\ \emph {et~al.}(2022)\citenamefont {Liu},
  \citenamefont {Chen}, \citenamefont {Zhang}, \citenamefont {Bockrath},
  \citenamefont {Lau}, \citenamefont {Zhou}, \citenamefont {Yoon},
  \citenamefont {Li}, \citenamefont {Liu}, \citenamefont {Dhale} \emph
  {et~al.}}]{Liu2022}%
  \BibitemOpen
  \bibfield  {author} {\bibinfo {author} {\bibfnamefont {Y.}~\bibnamefont
  {Liu}}, \bibinfo {author} {\bibfnamefont {R.}~\bibnamefont {Chen}}, \bibinfo
  {author} {\bibfnamefont {Z.}~\bibnamefont {Zhang}}, \bibinfo {author}
  {\bibfnamefont {M.}~\bibnamefont {Bockrath}}, \bibinfo {author}
  {\bibfnamefont {C.~N.}\ \bibnamefont {Lau}}, \bibinfo {author} {\bibfnamefont
  {Y.-F.}\ \bibnamefont {Zhou}}, \bibinfo {author} {\bibfnamefont
  {C.}~\bibnamefont {Yoon}}, \bibinfo {author} {\bibfnamefont {S.}~\bibnamefont
  {Li}}, \bibinfo {author} {\bibfnamefont {X.}~\bibnamefont {Liu}}, \bibinfo
  {author} {\bibfnamefont {N.}~\bibnamefont {Dhale}}, \emph {et~al.},\
  }\bibfield  {title} {\bibinfo {title} {Gate-{Tunable} {Transport} in
  {Quasi}-{One}-{Dimensional} $\alpha$-{Bi$_4$I$_4$} {Field} {Effect}
  {Transistors}},\ }\href {https://doi.org/10.1021/acs.nanolett.1c04264}
  {\bibfield  {journal} {\bibinfo  {journal} {Nano Lett.}\ }\textbf {\bibinfo
  {volume} {22}},\ \bibinfo {pages} {1151} (\bibinfo {year}
  {2022})}\BibitemShut {NoStop}%
\bibitem [{\citenamefont {Garate}(2013)}]{Garate2013}%
  \BibitemOpen
  \bibfield  {author} {\bibinfo {author} {\bibfnamefont {I.}~\bibnamefont
  {Garate}},\ }\bibfield  {title} {\bibinfo {title} {Phonon-{Induced}
  {Topological} {Transitions} and {Crossovers} in {Dirac} {Materials}},\ }\href
  {https://doi.org/10.1103/PhysRevLett.110.046402} {\bibfield  {journal}
  {\bibinfo  {journal} {Phys. Rev. Lett.}\ }\textbf {\bibinfo {volume} {110}},\
  \bibinfo {pages} {046402} (\bibinfo {year} {2013})}\BibitemShut {NoStop}%
\bibitem [{\citenamefont {Monserrat}\ and\ \citenamefont
  {Vanderbilt}(2016)}]{Monserrat2016}%
  \BibitemOpen
  \bibfield  {author} {\bibinfo {author} {\bibfnamefont {B.}~\bibnamefont
  {Monserrat}}\ and\ \bibinfo {author} {\bibfnamefont {D.}~\bibnamefont
  {Vanderbilt}},\ }\bibfield  {title} {\bibinfo {title} {Temperature {Effects}
  in the {Band} {Structure} of {Topological} {Insulators}},\ }\href
  {https://doi.org/10.1103/PhysRevLett.117.226801} {\bibfield  {journal}
  {\bibinfo  {journal} {Phys. Rev. Lett.}\ }\textbf {\bibinfo {volume} {117}},\
  \bibinfo {pages} {226801} (\bibinfo {year} {2016})}\BibitemShut {NoStop}%
\bibitem [{\citenamefont {Antonius}\ and\ \citenamefont
  {Louie}(2016)}]{Antonius2016}%
  \BibitemOpen
  \bibfield  {author} {\bibinfo {author} {\bibfnamefont {G.}~\bibnamefont
  {Antonius}}\ and\ \bibinfo {author} {\bibfnamefont {S.~G.}\ \bibnamefont
  {Louie}},\ }\bibfield  {title} {\bibinfo {title} {Temperature-{Induced}
  {Topological} {Phase} {Transitions}: {Promoted} versus {Suppressed}
  {Nontrivial} {Topology}},\ }\href
  {https://doi.org/10.1103/PhysRevLett.117.246401} {\bibfield  {journal}
  {\bibinfo  {journal} {Phys. Rev. Lett.}\ }\textbf {\bibinfo {volume} {117}},\
  \bibinfo {pages} {246401} (\bibinfo {year} {2016})}\BibitemShut {NoStop}%
\bibitem [{\citenamefont {Querales-Flores}\ \emph {et~al.}(2020)\citenamefont
  {Querales-Flores}, \citenamefont {Aguado-Puente}, \citenamefont
  {Dangi\ifmmode~\acute{c}\else \'{c}\fi{}}, \citenamefont {Cao}, \citenamefont
  {Chudzinski}, \citenamefont {Todorov}, \citenamefont {Gr\"uning},
  \citenamefont {Fahy},\ and\ \citenamefont {Savi\ifmmode~\acute{c}\else
  \'{c}\fi{}}}]{QueralesFlores2020}%
  \BibitemOpen
  \bibfield  {author} {\bibinfo {author} {\bibfnamefont {J.~D.}\ \bibnamefont
  {Querales-Flores}}, \bibinfo {author} {\bibfnamefont {P.}~\bibnamefont
  {Aguado-Puente}}, \bibinfo {author} {\bibfnamefont {D.}~\bibnamefont
  {Dangi\ifmmode~\acute{c}\else \'{c}\fi{}}}, \bibinfo {author} {\bibfnamefont
  {J.}~\bibnamefont {Cao}}, \bibinfo {author} {\bibfnamefont {P.}~\bibnamefont
  {Chudzinski}}, \bibinfo {author} {\bibfnamefont {T.~N.}\ \bibnamefont
  {Todorov}}, \bibinfo {author} {\bibfnamefont {M.}~\bibnamefont {Gr\"uning}},
  \bibinfo {author} {\bibfnamefont {S.}~\bibnamefont {Fahy}},\ and\ \bibinfo
  {author} {\bibfnamefont {I.}~\bibnamefont {Savi\ifmmode~\acute{c}\else
  \'{c}\fi{}}},\ }\bibfield  {title} {\bibinfo {title} {Towards
  temperature-induced topological phase transition in {SnTe}: A
  first-principles study},\ }\href
  {https://doi.org/10.1103/PhysRevB.101.235206} {\bibfield  {journal} {\bibinfo
   {journal} {Phys. Rev. B}\ }\textbf {\bibinfo {volume} {101}},\ \bibinfo
  {pages} {235206} (\bibinfo {year} {2020})}\BibitemShut {NoStop}%
\bibitem [{\citenamefont {Brousseau-Couture}\ \emph {et~al.}(2020)\citenamefont
  {Brousseau-Couture}, \citenamefont {Antonius},\ and\ \citenamefont
  {C\^ot\'e}}]{BrousseauCouture2020}%
  \BibitemOpen
  \bibfield  {author} {\bibinfo {author} {\bibfnamefont {V.}~\bibnamefont
  {Brousseau-Couture}}, \bibinfo {author} {\bibfnamefont {G.}~\bibnamefont
  {Antonius}},\ and\ \bibinfo {author} {\bibfnamefont {M.}~\bibnamefont
  {C\^ot\'e}},\ }\bibfield  {title} {\bibinfo {title} {Temperature dependence
  of the topological phase transition of {BiTeI} from first principles},\
  }\href {https://doi.org/10.1103/PhysRevResearch.2.023185} {\bibfield
  {journal} {\bibinfo  {journal} {Phys. Rev. Res.}\ }\textbf {\bibinfo {volume}
  {2}},\ \bibinfo {pages} {023185} (\bibinfo {year} {2020})}\BibitemShut
  {NoStop}%
\bibitem [{\citenamefont {Imai}\ \emph {et~al.}(2020)\citenamefont {Imai},
  \citenamefont {Chen}, \citenamefont {Kato}, \citenamefont {Kuroda},
  \citenamefont {Matsuda}, \citenamefont {Kimura}, \citenamefont {Miyamoto},
  \citenamefont {Eremeev},\ and\ \citenamefont {Okuda}}]{Imai2020}%
  \BibitemOpen
  \bibfield  {author} {\bibinfo {author} {\bibfnamefont {T.}~\bibnamefont
  {Imai}}, \bibinfo {author} {\bibfnamefont {J.}~\bibnamefont {Chen}}, \bibinfo
  {author} {\bibfnamefont {K.}~\bibnamefont {Kato}}, \bibinfo {author}
  {\bibfnamefont {K.}~\bibnamefont {Kuroda}}, \bibinfo {author} {\bibfnamefont
  {T.}~\bibnamefont {Matsuda}}, \bibinfo {author} {\bibfnamefont
  {A.}~\bibnamefont {Kimura}}, \bibinfo {author} {\bibfnamefont
  {K.}~\bibnamefont {Miyamoto}}, \bibinfo {author} {\bibfnamefont {S.~V.}\
  \bibnamefont {Eremeev}},\ and\ \bibinfo {author} {\bibfnamefont
  {T.}~\bibnamefont {Okuda}},\ }\bibfield  {title} {\bibinfo {title}
  {Experimental verification of a temperature-induced topological phase
  transition in {TlBiS}$_{2}$ and {TlBiSe}$_{2}$},\ }\href
  {https://doi.org/10.1103/PhysRevB.102.125151} {\bibfield  {journal} {\bibinfo
   {journal} {Phys. Rev. B}\ }\textbf {\bibinfo {volume} {102}},\ \bibinfo
  {pages} {125151} (\bibinfo {year} {2020})}\BibitemShut {NoStop}%
\bibitem [{\citenamefont {Giustino}\ \emph {et~al.}(2010)\citenamefont
  {Giustino}, \citenamefont {Louie},\ and\ \citenamefont
  {Cohen}}]{Giustino2010}%
  \BibitemOpen
  \bibfield  {author} {\bibinfo {author} {\bibfnamefont {F.}~\bibnamefont
  {Giustino}}, \bibinfo {author} {\bibfnamefont {S.~G.}\ \bibnamefont
  {Louie}},\ and\ \bibinfo {author} {\bibfnamefont {M.~L.}\ \bibnamefont
  {Cohen}},\ }\bibfield  {title} {\bibinfo {title} {{Electron-Phonon
  Renormalization of the Direct Band Gap of Diamond}},\ }\href
  {https://doi.org/10.1103/PhysRevLett.105.265501} {\bibfield  {journal}
  {\bibinfo  {journal} {Phys. Rev. Lett.}\ }\textbf {\bibinfo {volume} {105}},\
  \bibinfo {pages} {265501} (\bibinfo {year} {2010})}\BibitemShut {NoStop}%
\bibitem [{\citenamefont {Antonius}\ \emph {et~al.}(2014)\citenamefont
  {Antonius}, \citenamefont {Ponc\'e}, \citenamefont {Boulanger}, \citenamefont
  {C\^ot\'e},\ and\ \citenamefont {Gonze}}]{Antonius2014}%
  \BibitemOpen
  \bibfield  {author} {\bibinfo {author} {\bibfnamefont {G.}~\bibnamefont
  {Antonius}}, \bibinfo {author} {\bibfnamefont {S.}~\bibnamefont {Ponc\'e}},
  \bibinfo {author} {\bibfnamefont {P.}~\bibnamefont {Boulanger}}, \bibinfo
  {author} {\bibfnamefont {M.}~\bibnamefont {C\^ot\'e}},\ and\ \bibinfo
  {author} {\bibfnamefont {X.}~\bibnamefont {Gonze}},\ }\bibfield  {title}
  {\bibinfo {title} {{Many-Body Effects on the Zero-Point Renormalization of
  the Band Structure}},\ }\href
  {https://doi.org/10.1103/PhysRevLett.112.215501} {\bibfield  {journal}
  {\bibinfo  {journal} {Phys. Rev. Lett.}\ }\textbf {\bibinfo {volume} {112}},\
  \bibinfo {pages} {215501} (\bibinfo {year} {2014})}\BibitemShut {NoStop}%
\bibitem [{\citenamefont {Allen}\ and\ \citenamefont
  {Heine}(1976)}]{Allen1976}%
  \BibitemOpen
  \bibfield  {author} {\bibinfo {author} {\bibfnamefont {P.~B.}\ \bibnamefont
  {Allen}}\ and\ \bibinfo {author} {\bibfnamefont {V.}~\bibnamefont {Heine}},\
  }\bibfield  {title} {\bibinfo {title} {Theory of the temperature dependence
  of electronic band structures},\ }\href
  {https://doi.org/10.1088/0022-3719/9/12/013} {\bibfield  {journal} {\bibinfo
  {journal} {J. Phys. C: Solid State Phys.}\ }\textbf {\bibinfo {volume} {9}},\
  \bibinfo {pages} {2305} (\bibinfo {year} {1976})}\BibitemShut {NoStop}%
\bibitem [{\citenamefont {Allen}\ and\ \citenamefont
  {Cardona}(1981)}]{Allen1981}%
  \BibitemOpen
  \bibfield  {author} {\bibinfo {author} {\bibfnamefont {P.~B.}\ \bibnamefont
  {Allen}}\ and\ \bibinfo {author} {\bibfnamefont {M.}~\bibnamefont
  {Cardona}},\ }\bibfield  {title} {\bibinfo {title} {Theory of the temperature
  dependence of the direct gap of germanium},\ }\href
  {https://doi.org/10.1103/PhysRevB.23.1495} {\bibfield  {journal} {\bibinfo
  {journal} {Phys. Rev. B}\ }\textbf {\bibinfo {volume} {23}},\ \bibinfo
  {pages} {1495} (\bibinfo {year} {1981})}\BibitemShut {NoStop}%
\bibitem [{\citenamefont {Allen}\ and\ \citenamefont
  {Cardona}(1983)}]{Allen1983}%
  \BibitemOpen
  \bibfield  {author} {\bibinfo {author} {\bibfnamefont {P.~B.}\ \bibnamefont
  {Allen}}\ and\ \bibinfo {author} {\bibfnamefont {M.}~\bibnamefont
  {Cardona}},\ }\bibfield  {title} {\bibinfo {title} {Temperature dependence of
  the direct gap of {Si} and {Ge}},\ }\href
  {https://doi.org/10.1103/PhysRevB.27.4760} {\bibfield  {journal} {\bibinfo
  {journal} {Phys. Rev. B}\ }\textbf {\bibinfo {volume} {27}},\ \bibinfo
  {pages} {4760} (\bibinfo {year} {1983})}\BibitemShut {NoStop}%
\bibitem [{\citenamefont {Giustino}(2017)}]{Giustino2017}%
  \BibitemOpen
  \bibfield  {author} {\bibinfo {author} {\bibfnamefont {F.}~\bibnamefont
  {Giustino}},\ }\bibfield  {title} {\bibinfo {title} {Electron-phonon
  interactions from first principles},\ }\href
  {https://doi.org/10.1103/RevModPhys.89.015003} {\bibfield  {journal}
  {\bibinfo  {journal} {Rev. Mod. Phys.}\ }\textbf {\bibinfo {volume} {89}},\
  \bibinfo {pages} {015003} (\bibinfo {year} {2017})}\BibitemShut {NoStop}%
\bibitem [{\citenamefont {Nery}\ \emph {et~al.}(2018)\citenamefont {Nery},
  \citenamefont {Allen}, \citenamefont {Antonius}, \citenamefont {Reining},
  \citenamefont {Miglio},\ and\ \citenamefont {Gonze}}]{Nery2018}%
  \BibitemOpen
  \bibfield  {author} {\bibinfo {author} {\bibfnamefont {J.~P.}\ \bibnamefont
  {Nery}}, \bibinfo {author} {\bibfnamefont {P.~B.}\ \bibnamefont {Allen}},
  \bibinfo {author} {\bibfnamefont {G.}~\bibnamefont {Antonius}}, \bibinfo
  {author} {\bibfnamefont {L.}~\bibnamefont {Reining}}, \bibinfo {author}
  {\bibfnamefont {A.}~\bibnamefont {Miglio}},\ and\ \bibinfo {author}
  {\bibfnamefont {X.}~\bibnamefont {Gonze}},\ }\bibfield  {title} {\bibinfo
  {title} {Quasiparticles and phonon satellites in spectral functions of
  semiconductors and insulators: Cumulants applied to the full first-principles
  theory and the {Fr\"ohlich} polaron},\ }\href
  {https://doi.org/10.1103/PhysRevB.97.115145} {\bibfield  {journal} {\bibinfo
  {journal} {Phys. Rev. B}\ }\textbf {\bibinfo {volume} {97}},\ \bibinfo
  {pages} {115145} (\bibinfo {year} {2018})}\BibitemShut {NoStop}%
\bibitem [{\citenamefont {Ponc\'e}\ \emph {et~al.}(2014)\citenamefont
  {Ponc\'e}, \citenamefont {Antonius}, \citenamefont {Gillet}, \citenamefont
  {Boulanger}, \citenamefont {Laflamme~Janssen}, \citenamefont {Marini},
  \citenamefont {C\^ot\'e},\ and\ \citenamefont {Gonze}}]{Ponce2014}%
  \BibitemOpen
  \bibfield  {author} {\bibinfo {author} {\bibfnamefont {S.}~\bibnamefont
  {Ponc\'e}}, \bibinfo {author} {\bibfnamefont {G.}~\bibnamefont {Antonius}},
  \bibinfo {author} {\bibfnamefont {Y.}~\bibnamefont {Gillet}}, \bibinfo
  {author} {\bibfnamefont {P.}~\bibnamefont {Boulanger}}, \bibinfo {author}
  {\bibfnamefont {J.}~\bibnamefont {Laflamme~Janssen}}, \bibinfo {author}
  {\bibfnamefont {A.}~\bibnamefont {Marini}}, \bibinfo {author} {\bibfnamefont
  {M.}~\bibnamefont {C\^ot\'e}},\ and\ \bibinfo {author} {\bibfnamefont
  {X.}~\bibnamefont {Gonze}},\ }\bibfield  {title} {\bibinfo {title}
  {Temperature dependence of electronic eigenenergies in the adiabatic harmonic
  approximation},\ }\href {https://doi.org/10.1103/PhysRevB.90.214304}
  {\bibfield  {journal} {\bibinfo  {journal} {Phys. Rev. B}\ }\textbf {\bibinfo
  {volume} {90}},\ \bibinfo {pages} {214304} (\bibinfo {year}
  {2014})}\BibitemShut {NoStop}%
\bibitem [{\citenamefont {Lihm}\ and\ \citenamefont {Park}(2020)}]{Lihm2020}%
  \BibitemOpen
  \bibfield  {author} {\bibinfo {author} {\bibfnamefont {J.-M.}\ \bibnamefont
  {Lihm}}\ and\ \bibinfo {author} {\bibfnamefont {C.-H.}\ \bibnamefont
  {Park}},\ }\bibfield  {title} {\bibinfo {title} {Phonon-induced
  renormalization of electron wave functions},\ }\href
  {https://doi.org/10.1103/PhysRevB.101.121102} {\bibfield  {journal} {\bibinfo
   {journal} {Phys. Rev. B}\ }\textbf {\bibinfo {volume} {101}},\ \bibinfo
  {pages} {121102} (\bibinfo {year} {2020})}\BibitemShut {NoStop}%
\bibitem [{\citenamefont {van Schilfgaarde}\ \emph {et~al.}(2006)\citenamefont
  {van Schilfgaarde}, \citenamefont {Kotani},\ and\ \citenamefont
  {Faleev}}]{Schilfgaarde2006}%
  \BibitemOpen
  \bibfield  {author} {\bibinfo {author} {\bibfnamefont {M.}~\bibnamefont {van
  Schilfgaarde}}, \bibinfo {author} {\bibfnamefont {T.}~\bibnamefont
  {Kotani}},\ and\ \bibinfo {author} {\bibfnamefont {S.}~\bibnamefont
  {Faleev}},\ }\bibfield  {title} {\bibinfo {title} {Quasiparticle
  self-consistent {GW} theory},\ }\href
  {https://doi.org/10.1103/PhysRevLett.96.226402} {\bibfield  {journal}
  {\bibinfo  {journal} {Phys. Rev. Lett.}\ }\textbf {\bibinfo {volume} {96}},\
  \bibinfo {pages} {226402} (\bibinfo {year} {2006})}\BibitemShut {NoStop}%
\bibitem [{\citenamefont {Poncé}\ \emph {et~al.}(2025)\citenamefont {Poncé},
  \citenamefont {Lihm},\ and\ \citenamefont {Park}}]{Ponce2025}%
  \BibitemOpen
  \bibfield  {author} {\bibinfo {author} {\bibfnamefont {S.}~\bibnamefont
  {Poncé}}, \bibinfo {author} {\bibfnamefont {J.-M.}\ \bibnamefont {Lihm}},\
  and\ \bibinfo {author} {\bibfnamefont {C.-H.}\ \bibnamefont {Park}},\
  }\bibfield  {title} {\bibinfo {title} {Verification and validation of
  zero-point electron-phonon renormalization of the bandgap, mass enhancement,
  and spectral functions},\ }\href {https://doi.org/10.1038/s41524-025-01587-5}
  {\bibfield  {journal} {\bibinfo  {journal} {npj Comput. Mater.}\ }\textbf
  {\bibinfo {volume} {11}},\ \bibinfo {pages} {117} (\bibinfo {year}
  {2025})}\BibitemShut {NoStop}%
\bibitem [{\citenamefont {Gonze}\ \emph {et~al.}(2011)\citenamefont {Gonze},
  \citenamefont {Boulanger},\ and\ \citenamefont {Côté}}]{Gonze2011}%
  \BibitemOpen
  \bibfield  {author} {\bibinfo {author} {\bibfnamefont {X.}~\bibnamefont
  {Gonze}}, \bibinfo {author} {\bibfnamefont {P.}~\bibnamefont {Boulanger}},\
  and\ \bibinfo {author} {\bibfnamefont {M.}~\bibnamefont {Côté}},\
  }\bibfield  {title} {\bibinfo {title} {Theoretical approaches to the
  temperature and zero-point motion effects on the electronic band structure},\
  }\href {https://doi.org/https://doi.org/10.1002/andp.201000100} {\bibfield
  {journal} {\bibinfo  {journal} {Ann. Phys.}\ }\textbf {\bibinfo {volume}
  {523}},\ \bibinfo {pages} {168} (\bibinfo {year} {2011})}\BibitemShut
  {NoStop}%
\bibitem [{\citenamefont {Lihm}\ and\ \citenamefont {Park}(2021)}]{Lihm2021}%
  \BibitemOpen
  \bibfield  {author} {\bibinfo {author} {\bibfnamefont {J.-M.}\ \bibnamefont
  {Lihm}}\ and\ \bibinfo {author} {\bibfnamefont {C.-H.}\ \bibnamefont
  {Park}},\ }\bibfield  {title} {\bibinfo {title} {{Wannier} {Function}
  {Perturbation} {Theory}: {Localized} {Representation} and {Interpolation} of
  {Wave} {Function} {Perturbation}},\ }\href
  {https://doi.org/10.1103/PhysRevX.11.041053} {\bibfield  {journal} {\bibinfo
  {journal} {Phys. Rev. X}\ }\textbf {\bibinfo {volume} {11}},\ \bibinfo
  {pages} {041053} (\bibinfo {year} {2021})}\BibitemShut {NoStop}%
\bibitem [{sup()}]{supmat}%
  \BibitemOpen
  \href@noop {} {\bibinfo {title} {See {Supplemental Material} at link for
  details on the computational methods, convergence tests, surface states from
  {HSE} static-lattice calculations, calculated hinge states in
  {$\alpha$-Bi$_4$I$_4$}, comparison between calculated surface states and
  {ARPES} spectra, and temperature-dependent surface states in
  {$\beta$-Bi$_4$I$_4$}, which includes
  refs.~\cite{vanSetten2018,Perdew1996,Baroni2001,Kresse1996,Kresse1999,Heyd2003,Krukau2006,vonSchnering1978,Giustino2007,Sjakste2015,Verdi2015}}}\BibitemShut
  {NoStop}%
\bibitem [{\citenamefont {Giannozzi}\ \emph {et~al.}(2009)\citenamefont
  {Giannozzi}, \citenamefont {Baroni}, \citenamefont {Bonini}, \citenamefont
  {Calandra}, \citenamefont {Car} \emph {et~al.}}]{Giannozzi2009}%
  \BibitemOpen
  \bibfield  {author} {\bibinfo {author} {\bibfnamefont {P.}~\bibnamefont
  {Giannozzi}}, \bibinfo {author} {\bibfnamefont {S.}~\bibnamefont {Baroni}},
  \bibinfo {author} {\bibfnamefont {N.}~\bibnamefont {Bonini}}, \bibinfo
  {author} {\bibfnamefont {M.}~\bibnamefont {Calandra}}, \bibinfo {author}
  {\bibfnamefont {R.}~\bibnamefont {Car}}, \emph {et~al.},\ }\bibfield  {title}
  {\bibinfo {title} {{QUANTUM} {ESPRESSO}: a modular and open-source software
  project for quantum simulations of materials},\ }\href
  {https://doi.org/10.1088/0953-8984/21/39/395502} {\bibfield  {journal}
  {\bibinfo  {journal} {J. Phys.: Condens. Matter}\ }\textbf {\bibinfo {volume}
  {21}},\ \bibinfo {pages} {395502} (\bibinfo {year} {2009})}\BibitemShut
  {NoStop}%
\bibitem [{\citenamefont {Giannozzi}\ \emph {et~al.}(2017)\citenamefont
  {Giannozzi}, \citenamefont {Andreussi}, \citenamefont {Brumme}, \citenamefont
  {Bunau}, \citenamefont {Buongiorno~Nardelli} \emph {et~al.}}]{Giannozzi2017}%
  \BibitemOpen
  \bibfield  {author} {\bibinfo {author} {\bibfnamefont {P.}~\bibnamefont
  {Giannozzi}}, \bibinfo {author} {\bibfnamefont {O.}~\bibnamefont
  {Andreussi}}, \bibinfo {author} {\bibfnamefont {T.}~\bibnamefont {Brumme}},
  \bibinfo {author} {\bibfnamefont {O.}~\bibnamefont {Bunau}}, \bibinfo
  {author} {\bibfnamefont {M.}~\bibnamefont {Buongiorno~Nardelli}}, \emph
  {et~al.},\ }\bibfield  {title} {\bibinfo {title} {Advanced capabilities for
  materials modelling with {Quantum} {ESPRESSO}},\ }\href
  {https://doi.org/10.1088/1361-648X/aa8f79} {\bibfield  {journal} {\bibinfo
  {journal} {J. Phys. Condens. Matter}\ }\textbf {\bibinfo {volume} {29}},\
  \bibinfo {pages} {465901} (\bibinfo {year} {2017})}\BibitemShut {NoStop}%
\bibitem [{\citenamefont {Zhou}\ \emph {et~al.}(2021)\citenamefont {Zhou},
  \citenamefont {Park}, \citenamefont {Lu}, \citenamefont {Maliyov},
  \citenamefont {Tong},\ and\ \citenamefont {Bernardi}}]{Zhou2021}%
  \BibitemOpen
  \bibfield  {author} {\bibinfo {author} {\bibfnamefont {J.-J.}\ \bibnamefont
  {Zhou}}, \bibinfo {author} {\bibfnamefont {J.}~\bibnamefont {Park}}, \bibinfo
  {author} {\bibfnamefont {I.-T.}\ \bibnamefont {Lu}}, \bibinfo {author}
  {\bibfnamefont {I.}~\bibnamefont {Maliyov}}, \bibinfo {author} {\bibfnamefont
  {X.}~\bibnamefont {Tong}},\ and\ \bibinfo {author} {\bibfnamefont
  {M.}~\bibnamefont {Bernardi}},\ }\bibfield  {title} {\bibinfo {title}
  {Perturbo: {A} software package for ab initio electron–phonon interactions,
  charge transport and ultrafast dynamics},\ }\href
  {https://doi.org/10.1016/j.cpc.2021.107970} {\bibfield  {journal} {\bibinfo
  {journal} {Comput. Phys. Commun.}\ }\textbf {\bibinfo {volume} {264}},\
  \bibinfo {pages} {107970} (\bibinfo {year} {2021})}\BibitemShut {NoStop}%
\bibitem [{\citenamefont {Marzari}\ \emph {et~al.}(2012)\citenamefont
  {Marzari}, \citenamefont {Mostofi}, \citenamefont {Yates}, \citenamefont
  {Souza},\ and\ \citenamefont {Vanderbilt}}]{Marzari2012}%
  \BibitemOpen
  \bibfield  {author} {\bibinfo {author} {\bibfnamefont {N.}~\bibnamefont
  {Marzari}}, \bibinfo {author} {\bibfnamefont {A.~A.}\ \bibnamefont
  {Mostofi}}, \bibinfo {author} {\bibfnamefont {J.~R.}\ \bibnamefont {Yates}},
  \bibinfo {author} {\bibfnamefont {I.}~\bibnamefont {Souza}},\ and\ \bibinfo
  {author} {\bibfnamefont {D.}~\bibnamefont {Vanderbilt}},\ }\bibfield  {title}
  {\bibinfo {title} {Maximally localized wannier functions: Theory and
  applications},\ }\href {https://doi.org/10.1103/RevModPhys.84.1419}
  {\bibfield  {journal} {\bibinfo  {journal} {Rev. Mod. Phys.}\ }\textbf
  {\bibinfo {volume} {84}},\ \bibinfo {pages} {1419} (\bibinfo {year}
  {2012})}\BibitemShut {NoStop}%
\bibitem [{\citenamefont {Pizzi}\ \emph {et~al.}(2020)\citenamefont {Pizzi},
  \citenamefont {Vitale}, \citenamefont {Arita}, \citenamefont {Blügel},
  \citenamefont {Freimuth}, \citenamefont {G{\'{e}}ranton}, \citenamefont
  {Gibertini}, \citenamefont {Gresch}, \citenamefont {Johnson}, \citenamefont
  {Koretsune} \emph {et~al.}}]{Pizzi2020}%
  \BibitemOpen
  \bibfield  {author} {\bibinfo {author} {\bibfnamefont {G.}~\bibnamefont
  {Pizzi}}, \bibinfo {author} {\bibfnamefont {V.}~\bibnamefont {Vitale}},
  \bibinfo {author} {\bibfnamefont {R.}~\bibnamefont {Arita}}, \bibinfo
  {author} {\bibfnamefont {S.}~\bibnamefont {Blügel}}, \bibinfo {author}
  {\bibfnamefont {F.}~\bibnamefont {Freimuth}}, \bibinfo {author}
  {\bibfnamefont {G.}~\bibnamefont {G{\'{e}}ranton}}, \bibinfo {author}
  {\bibfnamefont {M.}~\bibnamefont {Gibertini}}, \bibinfo {author}
  {\bibfnamefont {D.}~\bibnamefont {Gresch}}, \bibinfo {author} {\bibfnamefont
  {C.}~\bibnamefont {Johnson}}, \bibinfo {author} {\bibfnamefont
  {T.}~\bibnamefont {Koretsune}}, \emph {et~al.},\ }\bibfield  {title}
  {\bibinfo {title} {Wannier90 as a community code: new features and
  applications},\ }\href {https://doi.org/10.1088/1361-648x/ab51ff} {\bibfield
  {journal} {\bibinfo  {journal} {J. Phys. Condens. Matter}\ }\textbf {\bibinfo
  {volume} {32}},\ \bibinfo {pages} {165902} (\bibinfo {year}
  {2020})}\BibitemShut {NoStop}%
\bibitem [{\citenamefont {Sancho}\ \emph {et~al.}(1985)\citenamefont {Sancho},
  \citenamefont {Sancho}, \citenamefont {Sancho},\ and\ \citenamefont
  {Rubio}}]{Sancho1985}%
  \BibitemOpen
  \bibfield  {author} {\bibinfo {author} {\bibfnamefont {M.~P.~L.}\
  \bibnamefont {Sancho}}, \bibinfo {author} {\bibfnamefont {J.~M.~L.}\
  \bibnamefont {Sancho}}, \bibinfo {author} {\bibfnamefont {J.~M.~L.}\
  \bibnamefont {Sancho}},\ and\ \bibinfo {author} {\bibfnamefont
  {J.}~\bibnamefont {Rubio}},\ }\bibfield  {title} {\bibinfo {title} {Highly
  convergent schemes for the calculation of bulk and surface {Green}
  functions},\ }\href {https://doi.org/10.1088/0305-4608/15/4/009} {\bibfield
  {journal} {\bibinfo  {journal} {J. Phys. F: Met. Phys.}\ }\textbf {\bibinfo
  {volume} {15}},\ \bibinfo {pages} {851} (\bibinfo {year} {1985})}\BibitemShut
  {NoStop}%
\bibitem [{\citenamefont {Li}\ \emph {et~al.}(2019)\citenamefont {Li},
  \citenamefont {Antonius}, \citenamefont {Wu}, \citenamefont {da~Jornada},\
  and\ \citenamefont {Louie}}]{Li2019}%
  \BibitemOpen
  \bibfield  {author} {\bibinfo {author} {\bibfnamefont {Z.}~\bibnamefont
  {Li}}, \bibinfo {author} {\bibfnamefont {G.}~\bibnamefont {Antonius}},
  \bibinfo {author} {\bibfnamefont {M.}~\bibnamefont {Wu}}, \bibinfo {author}
  {\bibfnamefont {F.~H.}\ \bibnamefont {da~Jornada}},\ and\ \bibinfo {author}
  {\bibfnamefont {S.~G.}\ \bibnamefont {Louie}},\ }\bibfield  {title} {\bibinfo
  {title} {{Electron-Phonon Coupling from Ab Initio Linear-Response Theory
  within the $GW$ Method: Correlation-Enhanced Interactions and
  Superconductivity in
  ${\mathrm{Ba}}_{1\ensuremath{-}x}{\mathrm{K}}_{x}{\mathrm{BiO}}_{3}$}},\
  }\href {https://doi.org/10.1103/PhysRevLett.122.186402} {\bibfield  {journal}
  {\bibinfo  {journal} {Phys. Rev. Lett.}\ }\textbf {\bibinfo {volume} {122}},\
  \bibinfo {pages} {186402} (\bibinfo {year} {2019})}\BibitemShut {NoStop}%
\bibitem [{\citenamefont {Abramovitch}\ \emph {et~al.}(2023)\citenamefont
  {Abramovitch}, \citenamefont {Zhou}, \citenamefont {Mravlje}, \citenamefont
  {Georges},\ and\ \citenamefont {Bernardi}}]{Abramovitch2023}%
  \BibitemOpen
  \bibfield  {author} {\bibinfo {author} {\bibfnamefont {D.~J.}\ \bibnamefont
  {Abramovitch}}, \bibinfo {author} {\bibfnamefont {J.-J.}\ \bibnamefont
  {Zhou}}, \bibinfo {author} {\bibfnamefont {J.}~\bibnamefont {Mravlje}},
  \bibinfo {author} {\bibfnamefont {A.}~\bibnamefont {Georges}},\ and\ \bibinfo
  {author} {\bibfnamefont {M.}~\bibnamefont {Bernardi}},\ }\bibfield  {title}
  {\bibinfo {title} {Combining electron-phonon and dynamical mean-field theory
  calculations of correlated materials: Transport in the correlated metal
  {Sr}$_{2}${RuO}$_{4}$},\ }\href
  {https://doi.org/10.1103/PhysRevMaterials.7.093801} {\bibfield  {journal}
  {\bibinfo  {journal} {Phys. Rev. Mater.}\ }\textbf {\bibinfo {volume} {7}},\
  \bibinfo {pages} {093801} (\bibinfo {year} {2023})}\BibitemShut {NoStop}%
\bibitem [{\citenamefont {Abramovitch}\ \emph {et~al.}(2024)\citenamefont
  {Abramovitch}, \citenamefont {Mravlje}, \citenamefont {Zhou}, \citenamefont
  {Georges},\ and\ \citenamefont {Bernardi}}]{Abramovitch2024}%
  \BibitemOpen
  \bibfield  {author} {\bibinfo {author} {\bibfnamefont {D.~J.}\ \bibnamefont
  {Abramovitch}}, \bibinfo {author} {\bibfnamefont {J.}~\bibnamefont
  {Mravlje}}, \bibinfo {author} {\bibfnamefont {J.-J.}\ \bibnamefont {Zhou}},
  \bibinfo {author} {\bibfnamefont {A.}~\bibnamefont {Georges}},\ and\ \bibinfo
  {author} {\bibfnamefont {M.}~\bibnamefont {Bernardi}},\ }\bibfield  {title}
  {\bibinfo {title} {{Respective Roles of Electron-Phonon and Electron-Electron
  Interactions in the Transport and Quasiparticle Properties of
  {SrVO}$_{3}$}},\ }\href {https://doi.org/10.1103/PhysRevLett.133.186501}
  {\bibfield  {journal} {\bibinfo  {journal} {Phys. Rev. Lett.}\ }\textbf
  {\bibinfo {volume} {133}},\ \bibinfo {pages} {186501} (\bibinfo {year}
  {2024})}\BibitemShut {NoStop}%
\bibitem [{\citenamefont {{van Setten}}\ \emph {et~al.}(2018)\citenamefont
  {{van Setten}}, \citenamefont {Giantomassi}, \citenamefont {Bousquet},
  \citenamefont {Verstraete}, \citenamefont {Hamann}, \citenamefont {Gonze},\
  and\ \citenamefont {Rignanese}}]{vanSetten2018}%
  \BibitemOpen
  \bibfield  {author} {\bibinfo {author} {\bibfnamefont {M.}~\bibnamefont {{van
  Setten}}}, \bibinfo {author} {\bibfnamefont {M.}~\bibnamefont {Giantomassi}},
  \bibinfo {author} {\bibfnamefont {E.}~\bibnamefont {Bousquet}}, \bibinfo
  {author} {\bibfnamefont {M.}~\bibnamefont {Verstraete}}, \bibinfo {author}
  {\bibfnamefont {D.}~\bibnamefont {Hamann}}, \bibinfo {author} {\bibfnamefont
  {X.}~\bibnamefont {Gonze}},\ and\ \bibinfo {author} {\bibfnamefont {G.-M.}\
  \bibnamefont {Rignanese}},\ }\bibfield  {title} {\bibinfo {title} {The
  {PseudoDojo}: {Training} and grading a 85 element optimized norm-conserving
  pseudopotential table},\ }\href
  {https://doi.org/https://doi.org/10.1016/j.cpc.2018.01.012} {\bibfield
  {journal} {\bibinfo  {journal} {Comput. Phys. Commun.}\ }\textbf {\bibinfo
  {volume} {226}},\ \bibinfo {pages} {39} (\bibinfo {year} {2018})}\BibitemShut
  {NoStop}%
\bibitem [{\citenamefont {Perdew}\ \emph {et~al.}(1996)\citenamefont {Perdew},
  \citenamefont {Burke},\ and\ \citenamefont {Ernzerhof}}]{Perdew1996}%
  \BibitemOpen
  \bibfield  {author} {\bibinfo {author} {\bibfnamefont {J.~P.}\ \bibnamefont
  {Perdew}}, \bibinfo {author} {\bibfnamefont {K.}~\bibnamefont {Burke}},\ and\
  \bibinfo {author} {\bibfnamefont {M.}~\bibnamefont {Ernzerhof}},\ }\bibfield
  {title} {\bibinfo {title} {Generalized {Gradient} {Approximation} {Made}
  {Simple}},\ }\href {https://doi.org/10.1103/PhysRevLett.77.3865} {\bibfield
  {journal} {\bibinfo  {journal} {Phys. Rev. Lett.}\ }\textbf {\bibinfo
  {volume} {77}},\ \bibinfo {pages} {3865} (\bibinfo {year}
  {1996})}\BibitemShut {NoStop}%
\bibitem [{\citenamefont {Baroni}\ \emph {et~al.}(2001)\citenamefont {Baroni},
  \citenamefont {de~Gironcoli}, \citenamefont {Dal~Corso},\ and\ \citenamefont
  {Giannozzi}}]{Baroni2001}%
  \BibitemOpen
  \bibfield  {author} {\bibinfo {author} {\bibfnamefont {S.}~\bibnamefont
  {Baroni}}, \bibinfo {author} {\bibfnamefont {S.}~\bibnamefont
  {de~Gironcoli}}, \bibinfo {author} {\bibfnamefont {A.}~\bibnamefont
  {Dal~Corso}},\ and\ \bibinfo {author} {\bibfnamefont {P.}~\bibnamefont
  {Giannozzi}},\ }\bibfield  {title} {\bibinfo {title} {Phonons and related
  crystal properties from density-functional perturbation theory},\ }\href
  {https://doi.org/10.1103/RevModPhys.73.515} {\bibfield  {journal} {\bibinfo
  {journal} {Rev. Mod. Phys.}\ }\textbf {\bibinfo {volume} {73}},\ \bibinfo
  {pages} {515} (\bibinfo {year} {2001})}\BibitemShut {NoStop}%
\bibitem [{\citenamefont {Kresse}\ and\ \citenamefont
  {Furthm\"uller}(1996)}]{Kresse1996}%
  \BibitemOpen
  \bibfield  {author} {\bibinfo {author} {\bibfnamefont {G.}~\bibnamefont
  {Kresse}}\ and\ \bibinfo {author} {\bibfnamefont {J.}~\bibnamefont
  {Furthm\"uller}},\ }\bibfield  {title} {\bibinfo {title} {Efficient iterative
  schemes for ab initio total-energy calculations using a plane-wave basis
  set},\ }\href {https://doi.org/10.1103/PhysRevB.54.11169} {\bibfield
  {journal} {\bibinfo  {journal} {Phys. Rev. B}\ }\textbf {\bibinfo {volume}
  {54}},\ \bibinfo {pages} {11169} (\bibinfo {year} {1996})}\BibitemShut
  {NoStop}%
\bibitem [{\citenamefont {Kresse}\ and\ \citenamefont
  {Joubert}(1999)}]{Kresse1999}%
  \BibitemOpen
  \bibfield  {author} {\bibinfo {author} {\bibfnamefont {G.}~\bibnamefont
  {Kresse}}\ and\ \bibinfo {author} {\bibfnamefont {D.}~\bibnamefont
  {Joubert}},\ }\bibfield  {title} {\bibinfo {title} {From ultrasoft
  pseudopotentials to the projector augmented-wave method},\ }\href
  {https://doi.org/10.1103/PhysRevB.59.1758} {\bibfield  {journal} {\bibinfo
  {journal} {Phys. Rev. B}\ }\textbf {\bibinfo {volume} {59}},\ \bibinfo
  {pages} {1758} (\bibinfo {year} {1999})}\BibitemShut {NoStop}%
\bibitem [{\citenamefont {Heyd}\ \emph {et~al.}(2003)\citenamefont {Heyd},
  \citenamefont {Scuseria},\ and\ \citenamefont {Ernzerhof}}]{Heyd2003}%
  \BibitemOpen
  \bibfield  {author} {\bibinfo {author} {\bibfnamefont {J.}~\bibnamefont
  {Heyd}}, \bibinfo {author} {\bibfnamefont {G.~E.}\ \bibnamefont {Scuseria}},\
  and\ \bibinfo {author} {\bibfnamefont {M.}~\bibnamefont {Ernzerhof}},\
  }\bibfield  {title} {\bibinfo {title} {Hybrid functionals based on a screened
  {Coulomb} potential},\ }\href
  {https://api.semanticscholar.org/CorpusID:55468013} {\bibfield  {journal}
  {\bibinfo  {journal} {J. Chem. Phys.}\ }\textbf {\bibinfo {volume} {118}},\
  \bibinfo {pages} {8207} (\bibinfo {year} {2003})}\BibitemShut {NoStop}%
\bibitem [{\citenamefont {Krukau}\ \emph {et~al.}(2006)\citenamefont {Krukau},
  \citenamefont {Vydrov}, \citenamefont {Izmaylov},\ and\ \citenamefont
  {Scuseria}}]{Krukau2006}%
  \BibitemOpen
  \bibfield  {author} {\bibinfo {author} {\bibfnamefont {A.~V.}\ \bibnamefont
  {Krukau}}, \bibinfo {author} {\bibfnamefont {O.~A.}\ \bibnamefont {Vydrov}},
  \bibinfo {author} {\bibfnamefont {A.~F.}\ \bibnamefont {Izmaylov}},\ and\
  \bibinfo {author} {\bibfnamefont {G.~E.}\ \bibnamefont {Scuseria}},\
  }\bibfield  {title} {\bibinfo {title} {Influence of the exchange screening
  parameter on the performance of screened hybrid functionals},\ }\href
  {https://doi.org/10.1063/1.2404663} {\bibfield  {journal} {\bibinfo
  {journal} {J. Chem. Phys.}\ }\textbf {\bibinfo {volume} {125}},\ \bibinfo
  {pages} {224106} (\bibinfo {year} {2006})}\BibitemShut {NoStop}%
\bibitem [{\citenamefont {von Schnering}\ \emph {et~al.}(1978)\citenamefont
  {von Schnering}, \citenamefont {von Benda},\ and\ \citenamefont
  {Kalveram}}]{vonSchnering1978}%
  \BibitemOpen
  \bibfield  {author} {\bibinfo {author} {\bibfnamefont {H.~G.}\ \bibnamefont
  {von Schnering}}, \bibinfo {author} {\bibfnamefont {H.}~\bibnamefont {von
  Benda}},\ and\ \bibinfo {author} {\bibfnamefont {C.}~\bibnamefont
  {Kalveram}},\ }\bibfield  {title} {\bibinfo {title} {Wismutmonojodid {BiJ},
  eine {Verbindung} mit {Bi(O)} und {Bi(II)}},\ }\href
  {https://doi.org/https://doi.org/10.1002/zaac.19784380104} {\bibfield
  {journal} {\bibinfo  {journal} {Zeitschrift für anorganische und allgemeine
  Chemie}\ }\textbf {\bibinfo {volume} {438}},\ \bibinfo {pages} {37} (\bibinfo
  {year} {1978})}\BibitemShut {NoStop}%
\bibitem [{\citenamefont {Giustino}\ \emph {et~al.}(2007)\citenamefont
  {Giustino}, \citenamefont {Cohen},\ and\ \citenamefont
  {Louie}}]{Giustino2007}%
  \BibitemOpen
  \bibfield  {author} {\bibinfo {author} {\bibfnamefont {F.}~\bibnamefont
  {Giustino}}, \bibinfo {author} {\bibfnamefont {M.~L.}\ \bibnamefont
  {Cohen}},\ and\ \bibinfo {author} {\bibfnamefont {S.~G.}\ \bibnamefont
  {Louie}},\ }\bibfield  {title} {\bibinfo {title} {Electron-phonon interaction
  using {Wannier} functions},\ }\href
  {https://doi.org/10.1103/PhysRevB.76.165108} {\bibfield  {journal} {\bibinfo
  {journal} {Phys. Rev. B}\ }\textbf {\bibinfo {volume} {76}},\ \bibinfo
  {pages} {165108} (\bibinfo {year} {2007})}\BibitemShut {NoStop}%
\bibitem [{\citenamefont {Sjakste}\ \emph {et~al.}(2015)\citenamefont
  {Sjakste}, \citenamefont {Vast}, \citenamefont {Calandra},\ and\
  \citenamefont {Mauri}}]{Sjakste2015}%
  \BibitemOpen
  \bibfield  {author} {\bibinfo {author} {\bibfnamefont {J.}~\bibnamefont
  {Sjakste}}, \bibinfo {author} {\bibfnamefont {N.}~\bibnamefont {Vast}},
  \bibinfo {author} {\bibfnamefont {M.}~\bibnamefont {Calandra}},\ and\
  \bibinfo {author} {\bibfnamefont {F.}~\bibnamefont {Mauri}},\ }\bibfield
  {title} {\bibinfo {title} {Wannier interpolation of the electron-phonon
  matrix elements in polar semiconductors: Polar-optical coupling in {GaAs}},\
  }\href {https://doi.org/10.1103/PhysRevB.92.054307} {\bibfield  {journal}
  {\bibinfo  {journal} {Phys. Rev. B}\ }\textbf {\bibinfo {volume} {92}},\
  \bibinfo {pages} {054307} (\bibinfo {year} {2015})}\BibitemShut {NoStop}%
\bibitem [{\citenamefont {Verdi}\ and\ \citenamefont
  {Giustino}(2015)}]{Verdi2015}%
  \BibitemOpen
  \bibfield  {author} {\bibinfo {author} {\bibfnamefont {C.}~\bibnamefont
  {Verdi}}\ and\ \bibinfo {author} {\bibfnamefont {F.}~\bibnamefont
  {Giustino}},\ }\bibfield  {title} {\bibinfo {title} {{Fr\"ohlich
  Electron-Phonon Vertex from First Principles}},\ }\href
  {https://doi.org/10.1103/PhysRevLett.115.176401} {\bibfield  {journal}
  {\bibinfo  {journal} {Phys. Rev. Lett.}\ }\textbf {\bibinfo {volume} {115}},\
  \bibinfo {pages} {176401} (\bibinfo {year} {2015})}\BibitemShut {NoStop}%
\end{thebibliography}%
\end{document}